\documentclass[preprint,amssymb,amsmath,bm,nofootinbib]{revtex4-1}
\usepackage{graphicx}

\def\simge{\mathrel{\rlap{\raise 0.511ex \hbox{$>$}}{\lower 0.511ex \hbox{$\sim$}}}}
\def\simle{\mathrel{\rlap{\raise 0.511ex \hbox{$<$}}{\lower 0.511ex \hbox{$\sim$}}}}
\def\slash#1{\setbox0=\hbox{$#1$}\dimen0=\wd0
     \setbox1=\hbox{/} \dimen1=\wd1 \ifdim\dimen0>\dimen1
     \rlap{\hbox to \dimen0{\hfil/\hfil}} #1                        \else
     \rlap{\hbox to \dimen1{\hfil$#1$\hfil}}
     /   \fi}

\def\slash#1{\mbox{$\not \!\! #1$}}

\def\lvec#1{\setbox0=\hbox{$#1$}
    \setbox1=\hbox{$\scriptstyle\leftarrow$}
    #1\kern-\wd0\smash{
    \raise\ht0\hbox{$\raise1pt\hbox{$\scriptstyle\leftarrow$}$}}
    \kern-\wd1\kern\wd0}
\def\rvec#1{\setbox0=\hbox{$#1$}
    \setbox1=\hbox{$\scriptstyle\rightarrow$}
    #1\kern-\wd0\smash{
    \raise\ht0\hbox{$\raise1pt\hbox{$\scriptstyle\rightarrow$}$}}
    \kern-\wd1\kern\wd0}

\def\diracstar#1#2{
    \setbox0=\hbox{$\gamma$}\setbox1=\hbox{$\gamma_{#1}$}
    \gamma_{#1}\kern-\wd1\kern\wd0
    \smash{\raise4.5pt\hbox{$\scriptstyle#2$}}}

\newcommand{\lsim}{
\mathrel{\hbox{\rlap{\hbox{\lower4pt\hbox{$\sim$}}}\hbox{$<$}}}}

\newcommand{\gsim}{
\mathrel{\hbox{\rlap{\hbox{\lower4pt\hbox{$\sim$}}}\hbox{$>$}}}}

\newcommand{\gev}{\, {\rm GeV}}
\newcommand{\mev}{\, {\rm MeV}}

\newcommand{\be}{\begin{equation}}
\newcommand{\ee}{\end{equation}}
\newcommand{\bea}{\begin{eqnarray}}
\newcommand{\eea}{\end{eqnarray}}

\begin{document}

\begin{flushright}
RM3-TH/16-2 \\
\end{flushright}


\centerline{\huge $K \to \pi$ semileptonic form factors} 
\vspace{0.5cm}
\centerline{\huge with $N_f=2+1+1$ Twisted Mass fermions}

\vspace{1.5cm}

\centerline{\large N.~Carrasco$^{(a)}$, P.~Lami$^{(b,a)}$, V.~Lubicz$^{(b,a)}$,} 
\vspace{0.25cm}
\centerline{\large L.~Riggio$^{(a)}$, S.~Simula$^{(a)}$, C.~Tarantino$^{(b,a)}$}

\vspace{1cm}

\centerline{\includegraphics[draft=false]{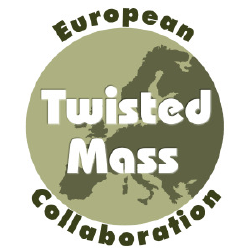}}

\vspace{1cm}

\centerline{\it $^{(a)}$ INFN, Sezione di Roma Tre}
\centerline{\it Via della Vasca Navale 84, I-00146 Rome, Italy}
\vskip 2 true mm
\centerline{\it $^{(b)}$ Dipartimento di Matematica e Fisica, Universit\`a degli Studi Roma Tre}
\centerline{\it Via della Vasca Navale 84, I-00146 Rome, Italy}

\begin{abstract}
We present a lattice QCD determination of the vector and scalar form factors of the semileptonic $K \to \pi \ell \nu$ decay which are relevant for the
extraction of the CKM matrix element $|V_{us}|$ from experimental data.
Our results are based on the gauge configurations produced by the European Twisted Mass Collaboration with $N_f = 2+1+1$ dynamical fermions, which include in the sea, besides two light mass degenerate quarks, also the strange and the charm quarks.
We use data simulated at three different values of the lattice spacing and with pion masses as small as $210$ MeV. 
Our final result for the vector form factor at zero momentum transfer is $f_+(0) = 0.9709 (46)$, where the uncertainty is both statistical and systematic combined in quadrature. 
Using the latest experimental value of $f_+(0) |V_{us}|$ from $K_{\ell 3}$ decays, we obtain $|V_{us}| = 0.2230 (11)$, which allows to test the unitarity constraint of the Standard Model below the permille level once the determination of $|V_{ud}|$ from superallowed nuclear $\beta$ decays is adopted.
A slight tension with unitarity at the level of $\sim 2$ standard deviations is observed.
Moreover we present our results for the semileptonic scalar $f_0(q^2)$ and vector $f_+(q^2)$ form factors in the whole range of values of the squared four-momentum transfer $q^2$ measured in $K_{\ell 3}$ decays, obtaining a very good agreement with the momentum dependence of the experimental data.
We provide a set of synthetic data points representing our results for the vector and scalar form factors at the physical point for several selected values of $q^2$.
\end{abstract}

\maketitle

\newpage

\section{Introduction}
\label{sec:intro}

In the Standard Model (SM) of particle physics the relative strength of the flavor-changing weak quark currents is parametrized by the unitary Cabibbo-Kobayashi-Maskawa (CKM) matrix \cite{CKM}, which provides the only source of CP violation in the quark sector.
The unitarity of the CKM matrix implies various relations between its matrix elements, known as the unitarity triangles.
The goal of flavor physics experiments is the measurement of the angles and sides of the unitarity triangles as well as the determination of many redundant observables sensitive to the CKM matrix elements.
The consistency of such a plethora of measurements would be spoiled by possible deviations from the SM, signalling the presence of New Physics (NP).
Therefore an accurate determination of the CKM matrix elements is crucial both for testing the SM and for searching NP. 

Let us focus on the determination of the Cabibbo's angle, or equivalently the CKM matrix element $|V_{us}|$, which enters both the leptonic kaon ($K_{\ell 2}$) decays via the axial quark current and the semileptonic kaon ($K_{\ell 3}$) decays via the vector quark current \cite{FlaviaNet}.
According to the $V - A$ structure of the SM weak current the decay rate of both processes should provide the same result for $|V_{us}|$, once the relevant hadronic quantities, namely the ratio of the kaon and pion leptonic decay constants $f_K / f_\pi$ and the semileptonic vector form factor at zero four-momentum transfer $f_+(0)$, are determined starting from our fundamental theory of the strong interactions, i.e.~from QCD.
As well known, such a task can be properly carried out by simulating QCD on a lattice.

The history of the efforts in predicting both $f_K / f_\pi$ and $f_+(0)$ using Lattice QCD (LQCD) is nicely summarised in the recent reviews \cite{FLAG1,FLAG2} of the Flavor Lattice Averaging Group (FLAG).
As far as the form factor $f_+(0)$ is concerned, the first attempt to calculate it on the lattice dates back to the work of Ref.~\cite{Becirevic:2004ya}, which was performed in the quenched approximation ($N_f = 0$) with simulated pions not lighter than $\simeq 500$ MeV.
Since then several improvements have been achieved and the most important ones are the inclusion of the effects of the loops of dynamical sea quarks, thanks to the production of unquenched gauge ensembles with $N_f = 2$, $2+1$ and $2+1+1$ sea quarks, and the simulation of lighter pions, which have recently reached the physical point \cite{Bazavov:2013maa,Boyle:2015hfa}.
Only one calculation of $f_+(0)$ exists to date \cite{Bazavov:2013maa} with four flavors of dynamical quarks ($N_f = 2+1+1$), which include in the sea, besides two light mass degenerate quarks, also the strange and the charm quarks.

In this work we present a new LQCD prediction of the form factor $f_+(0)$ obtained using the ensembles of gauge configurations produced by the European Twisted Mass (ETM) Collaboration with $N_f = 2+1+1$ dynamical quarks \cite{Baron:2010bv,Baron:2011sf}.
Furthermore, we have evaluated the strangeness changing vector $f_+(q^2)$ and scalar $f_0(q^2)$ form factors in the whole range of values of the squared four-momentum transfer $q^2$ measured in $K_{\ell 3}$ decays, obtaining a very good agreement with the momentum dependence of the experimental data \cite{FlaviaNet,Moulson:2014cra}.
Preliminary results have been already presented in Refs.~\cite{Carrasco:2014pta,Carrasco:2014uda,Carrasco:2015wzu}.

The gauge ensembles and the simulations used in this work are the same adopted in Ref.~\cite{Carrasco:2014cwa} to determine the up, down, strange and charm quark masses (see Tables 1-3 of Ref.~\cite{Carrasco:2014cwa}), using the experimental value of the pion decay constant, $f_\pi$, to set the lattice scale\footnote{With respect to Ref.~\cite{Carrasco:2014cwa} the number of independent gauge configurations adopted for the ensemble D15.48 has been increased to $90$ to improve the statistics.}, as well as in Ref.~\cite{Carrasco:2014poa} to calculate the ratio of leptonic decay constants $f_K / f_\pi$. 
The gauge fields are simulated using the Iwasaki gluon action \cite{Iwasaki:1985we}, while sea quarks are implemented with the Wilson Twisted Mass Action at maximal twist \cite{Frezzotti:2003xj, Frezzotti:2003ni}. 
In order to avoid the mixing of strange and charm quarks in the valence sector we have adopted the non-unitary setup described in Ref.~\cite{Frezzotti:2004wz}, in which the valence strange quarks are regularized as Osterwalder-Seiler (OS) fermions \cite{Osterwalder:1977pc}, while the valence up and down quarks have the same action of the sea.
The strange and charm mixing can occur in the hadronic decomposition of correlation functions only when the valence strange and charm quarks are regularised using the same twisted-mass doublet adopted for the strange and charm sea quarks (see Eq.~(8) of Ref.~\cite{Carrasco:2014cwa}). The use of different lattice regularisations for valence and sea heavy quarks avoids completely the effects of flavor mixing without modifying the renormalization pattern of operators in massless schemes and produces only a modification of discretization effects.
Moreover, since we work at maximal twist (i.e.~we impose that the bare Wilson mass is tuned at its critical value, which is common to all the different quarks), an automatic ${\cal{O}}(a)$-improvement \cite{Frezzotti:2003ni, Frezzotti:2004wz} is guaranteed also for our non-unitary setup.

The simulations have been carried out at three different values of the inverse bare lattice coupling $\beta$ to allow for a controlled extrapolation to the continuum limit, and different lattice volumes.
For each gauge ensemble we have used a number of gauge configurations corresponding to a separation of 20 trajectories to avoid autocorrelations.  
For the light sector we have simulated quark masses in the range from $3 m_{ud}$ to $12 m_{ud}$ and for the strange sector from $0.7 m_s$ to $1.2 m_s$, where $m_{ud}$ and $m_s$ are the physical values of the (renormalized) average up/down and strange quark masses, respectively, as determined in Ref.~\cite{Carrasco:2014cwa}.
The lattice spacings are found to be $a = \{ 0.0885(36), 0.0815(30), 0.0619(18) \}$ fm at $\beta = \{1.90, 1.95, 2.10\}$ respectively, the lattice volume goes from $\simeq 2$ to $\simeq 3$ fm, and the pion masses, extrapolated to the continuum and infinite volume limits, range from $\simeq 210$ to $ \simeq 450 \mev$ (see later Table~\ref{tab:masses&simudetails} and Ref.~\cite{Carrasco:2014cwa} for further details).

We present our study of the semileptonic $K \to \pi$ form factors using the results of the eight branches of the analysis carried out in Ref.~\cite{Carrasco:2014cwa} for determining the up, down, strange and charm quark masses. 
The various branches differ by: ~ i) the choice of the scaling variable, which was taken to be either the Sommer parameter $r_0/a$ \cite{Sommer:1993ce} or the mass of a fictitious pseudoscalar (PS) meson made of two strange-like quarks $a M_{s^\prime s^\prime}$; ~ ii) the fitting procedures, which were based either on Chiral Perturbation Theory (ChPT) or on a polynomial expansion in the light quark mass (for the motivations see the discussion in Section 3.1 of Ref.~\cite{Carrasco:2014cwa}); and ~ iii)  the choice between two methods, denoted as M1 and M2 which differ by ${\cal{O}}(a^2)$ effects (see, e.g., Ref.~\cite{Constantinou:2010gr}), used to determine non-perturbatively the values of the mass renormalization constant (RC) $Z_m = 1 / Z_P$~\cite{Carrasco:2014cwa}.
Throughout this work the results corresponding to the various branches of the analysis are combined to form our average and error according to Eq.~(28) of Ref.~\cite{Carrasco:2014cwa}.

The final results for the form factor $f_+(0)$ and the CKM matrix element $|V_{us}|$ we present in this paper are
 \be
     f_+(0) = 0.9709 ~ (46) ~ , \qquad |V_{us}| = 0.2230 ~ (11) ~ .
     \label{eq:results}
 \ee
Once the determinations of $|V_{ud}|$ from superallowed nuclear $\beta$ decays \cite{Hardy:2014qxa} and of $|V_{ub}|$ from $B$-meson decays \cite{PDG} are adopted, the unitarity test of the first-row of the CKM matrix becomes
 \be
      |V_{ud}|^2 + |V_{us}|^2 + |V_{ub}|^2 = 0.99875 ~ (41)_{V_{ud}} ~ (49)_{V_{us}} = 0.99875 ~ (64) ~ ,
      \label{eq:unitarity_Kl3}
 \ee
which highlights a slight tension with unitarity at the level of $\sim 2$ standard deviations.
Note also that the total uncertainty in the r.h.s.~of Eq.~(\ref{eq:unitarity_Kl3}) is below the permille level and the contribution due to the error on $|V_{us}|$ is comparable with the one related to $|V_{ud}|$.

As for the $q^2$-dependence of the semileptonic vector $f_+(q^2)$ and scalar $f_0(q^2)$ form factors, the same dispersive parameterization used to describe the experimental data \cite{FlaviaNet,Moulson:2014cra} has been adopted to fit our results extrapolated to the physical point. 
The dispersive fit depends on two parameters, $\Lambda_+$ and $C$, which represent respectively the slope of the vector form factor $f_+(q^2)$ at $q^2 = 0$ (in units of $M_\pi^2$) and the scalar form factor $f_0(q^2)$ at the (unphysical) Callan-Treiman (CT) point $q^2 = q_{CT}^2 \equiv M_K^2 - M_\pi^2$ \cite{Callan:1966hu} divided by $f_+(0)$.
Our final results are
 \be
     \Lambda_+ = 24.22 ~ (1.16) \cdot 10^{-3} ~, \qquad \rm{log}(C) = 0.1998 ~ (138) ~ ,
     \label{eq:results_dispersive}
 \ee
which compare positively with the latest experimental results \cite{Moulson:2014cra}
 \be
     \Lambda_+^{exp} = 25.75 ~ (36) \cdot 10^{-3} ~, \qquad \rm{log}(C)^{exp} = 0.1985 ~ (70) ~ .
     \label{eq:exp_results_dispersive}
 \ee

The physics described by our simulations and used throughout this work corresponds to the isospin symmetric limit of QCD, where $m_u = m_d = m_{ud}$, assuming also zero quark electric charges.
Therefore, isospin breaking and electromagnetic corrections have to be added separately in phenomenological analyses (see, e.g., \cite{FlaviaNet}).

The paper is organized as follows.
In Section \ref{sec:computation} we present the computational strategies adopted to extract the semileptonic form factors $f_+(q^2)$ and $f_0(q^2)$ from the 2- and 3-point correlation functions evaluated on the lattice using both the vector and scalar strangeness changing quark currents.
In Section \ref{sec:first_strategy} and \ref{sec:second_strategy} we present our results at the physical point and in the continuum limit coming from two different strategies.
The first one is based on the study of the momentum dependence of the form factors $f_+(q^2)$ and $f_0(q^2)$ to obtain the form factor $f_+(0)$ at each simulated light-quark mass $m_\ell$ and then to extrapolate the results for $f_+(0)$ to the physical point $m_\ell = m_{ud}$ and to the continuum limit.
The second strategy uses a combined fit of the $q^2$, $m_{\ell}$ and lattice spacing dependencies of our form factor data, obtaining in this way the momentum dependence of the form factors at the physical point for both spacelike and timelike $q^2$.
We provide a set of synthetic data points representing our results for $f_+(q^2)$ and $f_0(q^2)$ at the physical point for several selected values of $q^2$, including also the covariance matrix for the data at different values of $q^2$.
We compare our results with the experimental data obtaining a very good agreement in the whole range of values of $q^2$ measured in $K_{\ell 3}$ decays.
In Section \ref{sec:Vus} we present our determination of the CKM matrix element $|V_{us}|$ and we test the unitarity of the first-row of the CKM matrix.
Our conclusions are summarised in Section \ref{sec:conclusions}.

\section{Lattice computation of the form factors}
\label{sec:computation}

The matrix element of the strangeness changing vector current $\hat{V}^\mu$ between kaon and pion states decomposes into two form factors, $f_+$ and $f_-$, as
 \be
     \label{eq:matrixelement_Vmu}
     \langle \hat{V}_\mu \rangle  \equiv \langle  \pi(p_\pi) | \hat{V}_{\mu} | K(p_K) \rangle = (p_K + p_\pi)_\mu ~ f_+(q^2) + (p_K - p_\pi)_\mu ~ f_-(q^2) ~ ,
 \ee
where $q_\mu = (p_K - p_\pi)_\mu$ is the $4-$momentum transferred between the kaon and the pion.

The scalar form factor $f_0$ is defined as
 \be
     \label{eq:f0def}
     f_0(q^2) \equiv f_+(q^2)  + \frac{q^2}{M_K^2 - M_\pi^2} f_-(q^2)
 \ee
and satisfies by definition the relation $f_+(0) = f_0(0)$.
The form factor $f_0$ is proportional to the 4-divergence of $\langle \hat{V}_\mu \rangle$, which in turn, thanks to vector Ward-Takahashi identity, is related to the matrix element of the strangeness changing scalar density $\hat{S}$ between kaon and pion states, leading to
 \be
      \label{eq:matrixelement_S}
      \langle \hat{S} \rangle \equiv \langle  \pi(p_\pi) | \hat{S} | K(p_K)  \rangle = \frac{M_K^2 - M_\pi^2}{m_s - m_{ud}} f_0(q^2) ~ .
 \ee

Equations (\ref{eq:matrixelement_Vmu}) and (\ref{eq:matrixelement_S}) represent a system of redundant relations between the two form factors $f_+(q^2)$ and $f_0(q^2)$ and the matrix elements $\langle \hat{V}_\mu \rangle$ and $\langle  \hat{S} \rangle$.
The latter can be determined by studying the time dependence of suitable combinations of (Euclidean) 2- and 3-point correlation functions, namely
 \bea
      \label{eq:C2}
      C_2^{\pi(K)}(t^\prime; \vec{p}_{\pi(K)}) & = & \frac{1}{L^3} \sum\limits_{\vec{x}, \vec{z}} \left\langle 0 \right| P_5^{\pi(K)}(x) P_5^{\pi(K) \dag}(z) \left| 0 \right\rangle 
                                                                             e^{-i \vec{p}_{\pi(K)} \cdot (\vec{x} - \vec{z})} ~ \delta_{t^\prime, (t_x  - t_z )} , \\
      \label{eq:C3}
      C_{\hat{\Gamma}}^{K \pi}(t, t^\prime; \vec{p}_K, \vec{p}_\pi) & = & \frac{1}{L^6} \sum\limits_{\vec{x}, \vec{y}, \vec{z}} \langle 0 | P_5^\pi (x) \hat{\Gamma} (y) P_5^{K \dag} (z)  
                                                 | 0 \rangle e^{-i \vec{p}_K \cdot (\vec{y} - \vec{z}) + i \vec{p}_\pi \cdot (\vec{y} - \vec{x})} ~  \delta_{t, (t_y  - t_z )} ~ \delta_{t^\prime, (t_x  - t_z )} , ~
 \eea
where  $t^\prime$ is the time distance between the source and the sink, $t$ is the time distance between the insertion of the current $\Gamma \in \{\hat{V}_\mu, \hat{S}\}$ and the source, $P_5^\pi(x) = i \overline{u}(x) \gamma_5 d(x)$ and $P_5^K(x) = i \overline{s}(x) \gamma_5 d(x)$ are the (local) interpolating fields of the pion and kaon mesons.
In our lattice setup the Wilson parameters of the two valence quarks in any PS meson are always chosen to have opposite values, i.e.~$r_s = r_u = -r_d$.
In this way the squared PS mass differs from its continuum counterpart only by terms of ${\cal{O}}(a^2 m)$ \cite{Frezzotti:2003ni}.

The statistical accuracy of the correlators (\ref{eq:C2}-\ref{eq:C3}) can be significantly improved by using the all-to-all quark propagators evaluated with the so-called ``one-end" stochastic method \cite{McNeile:2006bz}, which includes spatial stochastic sources at a single time slice chosen randomly (see Ref.~\cite{Frezzotti:2008dr}, where the degenerate case of the pion is illustrated in details).
Statistical errors are always evaluated using the jackknife procedure, while cross-correlations are taken into account by the use of the eight bootstrap samplings corresponding to the eight analyses of Ref.~\cite{Carrasco:2014cwa} summarised in Section \ref{sec:intro}.
In Table \ref{tab:masses&simudetails} we summarize the basic simulation parameters as well as the pion and kaon masses corresponding to each ensemble used in this work.

\begin{table}[hbt!]
\begin{center}
\renewcommand{\arraystretch}{1.20}
\begin{tabular}{||c|c|c||c|c|c|c||c|c|c|c||}
\hline
ensemble & $\beta$ & $V / a^4$ &$a\mu_{sea}=a\mu_\ell$&$a\mu_\sigma$&$a\mu_\delta$& $a\mu_s$& $M_\pi {\rm (MeV)}$ & $M_K {\rm (MeV)}$ & $L {\rm (fm)}$ & $M_\pi L$\\
\hline \hline
$A30.32$ & $1.90$ & $32^3 \times 64$ &$0.0030$ & $0.15$ & $0.19$ & $\{0.0145,$& 275 & 577 & 2.84 & 3.96 \\
$A40.32$ & & & $0.0040$ & & & $0.0185,$ & 315 & 588 & & 4.53 \\
$A50.32$ & & & $0.0050$ & & & $0.0225\}$ & 350 & 595 & & 5.04 \\
\cline{1-1} \cline{3-4} \cline{8-11} 
$A40.24$ & & $24^3 \times 48 $ & $0.0040$ & & & & 324 & 594 & 2.13 & 3.50 \\
$A60.24$ & & & $0.0060$ & & & & 388 & 610 & & 4.19 \\
$A80.24$ & & & $0.0080$ & & & & 438 & 624 & & 4.73 \\
$A100.24$ & & & $0.0100$ & & & & 497 & 650 & & 5.37 \\
\hline \hline
$B25.32$ & $1.95$ & $32^3 \times 64$ & $0.0025$ & $0.135$ & $0.170$ & $\{0.0141,$& 259 & 553 & 2.61 & 3.43 \\
$B35.32$ & & & $0.0035$ & & & $ 0.0180,$ & 300 & 562 & & 3.97 \\
$B55.32$ & & & $0.0055$ & & & $0.0219\}$ & 377 & 587 & & 4.99 \\
$B75.32$ &  & & $0.0075$ & & & & 437 & 608 & & 5.78 \\
\cline{1-1} \cline{3-4} \cline{8-11} 
$B85.24$ & & $24^3 \times 48 $ & $0.0085$ & & & & 463 & 617 & 1.96 & 4.60 \\
\hline \hline
$D15.48$ & $2.10$ & $48^3 \times 96$ & $0.0015$ & $0.12$ &$0.1385 $& $\{0.0118,$ & 224 & 538 & 2.97 & 3.37 \\ 
$D20.48$ & & & $0.0020$ & & & $0.0151,$ & 255 & 541 & & 3.84 \\
$D30.48$ & & & $0.0030$ & & & $0.0184\}$ & 310 & 554 & & 4.67 \\
 \hline   
\end{tabular}
\renewcommand{\arraystretch}{1.0}
\end{center}
\caption{\it Values of the simulated sea and valence quark bare masses, of the pion ($M_\pi$) and kaon ($M_K$) masses, of the lattice size $L$ and of the product $M_\pi L$ for the various gauge ensembles used in this work. The values of $M_K$ given for each gauge ensemble do not correspond to the value of the simulated strange bare mass shown in the seventh column, but to a renormalized strange mass interpolated at the physical value $m_s = 99.6 (4.3)$ MeV \cite{Carrasco:2014cwa}.}
\label{tab:masses&simudetails}
\end{table}

At large values of the time distances $t$, $t^\prime$ and ($t^\prime - t$) one has
 \bea
        \label{eq:C2_larget}
        C_2^{\pi(K)}(t^\prime; \vec{p}_{\pi(K)}) & ~ _{\overrightarrow{t^\prime \gg a}} ~ & \frac{|Z_{\pi(K)}|^2}{2E_{\pi(K)}} \left[ e^{-E_{\pi(K)} t^\prime} + 
                                           e^{-E_{\pi(K)} (T - t^\prime)} \right] , \\
        \label{eq:C3_larget}        
        C_{\hat{\Gamma}}^{K \pi}(t, t^\prime; \vec{p}_K, \vec{p}_\pi) & ~ _{\overrightarrow{t, t^\prime, (t^\prime - t) \gg a}} ~ &  \frac{Z_\pi Z_K^*}{4E_\pi E_K} ~
                                            \langle  \pi(p_\pi) | \hat{\Gamma} | K(p_K) \rangle ~ e^{-E_K t} ~ e^{-E_\pi (t^\prime - t)}
 \eea
where $E_{\pi(K)}$ is the pion (kaon) energy and $Z_{\pi(K)} \equiv \langle 0| P_5^{\pi(K)}(0) | \pi(K) \rangle$ is independent on the pion (kaon) momentum up to discretization effects.
Using the exponential fit given by the r.h.s~of Eq.~(\ref{eq:C2_larget}), the pion and kaon energies can be extracted directly from the corresponding 2-point correlation functions.
The time intervals $[t_{min}, t_{max}]$ adopted for the fit (\ref{eq:C2_larget}) can be read off from Table 4 of Ref.~\cite{Carrasco:2014cwa}. 
We have checked that: ~ i) the changes in the meson energies due to a decrease in the value of $t_{min}$ by one or two lattice units are well below the statistical uncertainty, and ~ ii) the extracted values of $E_{\pi(K)}$ are nicely reproduced (within statistical errors) by the continuum-like dispersive relation $E_{\pi(K)}^{disp.} = \sqrt{M_{\pi(K)}^2 + |\vec{p}_{\pi(K)}|^2}$, where $M_{\pi(K)}$ is the meson mass extracted from the 2-point correlator corresponding to the meson at rest.

In our lattice setup we employs maximally twisted fermions and therefore the vector $\hat{V}_\mu$ and the scalar $\hat{S}$ currents renormalize multiplicatively \cite{Frezzotti:2003ni}.
Introducing the local bare currents $V_\mu(x) \equiv \overline{s}(x) \gamma_\mu u(x)$ and $S(x) \equiv \overline{s}(x) u(x)$ and keeping the same value for the Wilson parameters of the initial and final quarks (i.e.~$r_s = r_u$), one has
 \bea
       \label{eq:Vmu}
        \hat{V}_\mu(x) & = & Z_V ~ V_\mu(x) = Z_V ~ \overline{s}(x) \gamma_\mu u(x) ~ , \\
       \label{eq:S}
        \hat{S}(x) & = & Z_P ~ S(x) = Z_P ~ \overline{s}(x) u(x) ~ ,
 \eea
where $Z_V$ and $Z_P$ are the renormalization constants (RCs) of the vector and pseudoscalar densities for standard Wilson fermions, respectively, determined using the RI'-MOM method in Ref.~\cite{Carrasco:2014cwa}.
The twisted quark masses renormalize multiplicatively with a RC $Z_m$ given by $Z_m = 1 / Z_P$ \cite{Frezzotti:2003ni}, which means that the product $(m_s - m_{ud}) \langle \hat{S} \rangle$ does not depend on the RC $Z_P$.
Therefore, according to Eq.~(\ref{eq:matrixelement_S}), the scalar form factor $f_0(q^2)$ is related to the (bare) matrix element $\langle S \rangle$ by
 \be
      \label{eq:f0_S}
      \langle S \rangle \equiv \langle  \pi(p_\pi) | S | K(p_K) \rangle = \frac{M_K^2 - M_\pi^2}{\mu_s - \mu_\ell} f_0(q^2) ~ ,
 \ee
where $\mu_\ell$ and $\mu_s$ are the light and strange bare quark masses, respectively. 

The (renormalized) matrix elements $\langle \hat{V}_\mu \rangle$ can be extracted from the time dependence of suitable ratios $R_\mu^{(V)}$ of the 3-point correlation functions, defined as in Eq.~(\ref{eq:C3}) but using the local bare current $V_\mu(x)$, namely for $\mu = 0, 1, 2, 3$
 \be
      \label{eq:RVmu}
      R_\mu^{(V)}(t; \vec{p}_K, \vec{p}_\pi) \equiv 4 p_{K \mu} p_{\pi \mu} ~ \frac{C_{V_\mu}^{K \pi}(t, \frac{T}{2}; \vec{p}_K, \vec{p}_\pi) ~ 
      C_{V_\mu}^{\pi K}(t, \frac{T}{2};  \vec{p}_\pi, \vec{p}_K)}{C_{V_\mu}^{\pi \pi}(t, \frac{T}{2};  \vec{p}_\pi, \vec{p}_\pi) ~ 
      C_{V_\mu}^{K K}(t, \frac{T}{2}; \vec{p}_K, \vec{p}_K)} ~ ,
 \ee
where the time distance between the source and the sink has been fixed\footnote{This choice allows to improve the statistics by averaging the correlation functions between the two halves of the time extension of the lattice.} at $t^\prime = T/2$. 
We point out that the denominator of Eq.~(\ref{eq:RVmu}) is simply the numerator evaluated in the mass-degenerate limit for the valence quarks in the current. 
Denoting the bare masses of the quarks in the current as ($\mu_1, \mu_2$), the two correlation functions in the numerator correspond to ($\mu_s, \mu_\ell$) and ($\mu_\ell, \mu_s$), respectively, while the denominator refers to the cases ($\mu_\ell, \mu_\ell$) and ($\mu_s, \mu_s$). 
Though mass-degenerate the current quarks in the denominator have the same lattice regularisations of the corresponding quarks in the numerator. 
Thus the same RC $Z_V$ is present at numerator and denominator and it cancels out in the ratio.

At large time distances $t$ a plateau is expected to occur, viz.
 \be
    \label{eq:RVmu_larget}
    R_\mu^{(V)}(t; \vec{p}_K, \vec{p}_\pi) _{ ~ \overrightarrow{t \gg a, ~ (T/2 - t) \gg a} ~ } 4 p_{K \mu} p_{\pi \mu} ~ \frac{\langle \pi(p_\pi) | \hat{V}_\mu | K(p_K) \rangle 
    \langle K(p_K) | \hat{V}_\mu| \pi(p_\pi) \rangle}{\langle  \pi(p_\pi) | \hat{V}_\mu| \pi(p_\pi) \rangle \langle K(p_K) | \hat{V}_\mu | K(p_K) \rangle} ~ ,
 \ee
which, we stress, does not depend on the RC $Z_V$ and on the matrix elements $Z_\pi$ and $Z_K$ of the interpolating PS fields.
Moreover, thanks to charge conservation one has $\langle  \pi(p_\pi) | \hat{V}_\mu| \pi(p_\pi) \rangle = 2 p_{\pi \mu}$ and $\langle  K(p_K) | \hat{V}_\mu| K(p_K) \rangle = 2 p_{K \mu}$ up to lattice artefacts, and therefore the plateau of $R_\mu^{(V)}(t; \vec{p}_K, \vec{p}_\pi)$ at large time distances provides directly the quantity $|\langle \hat{V}_\mu \rangle|^2$.
Taking the square root the absolute value of the matrix element $\langle \hat{V}_\mu \rangle$ is obtained, while the sign of $\langle \hat{V}_\mu \rangle$ can be easily inferred from the sign of the plateau of the ratio $C_{V_\mu}^{K \pi}(t, \frac{T}{2}; \vec{p}_K, \vec{p}_\pi) /$ $[ C_2^K(t; \vec{p}_K) ~ C_2^\pi(T/2 - t; \vec{p}_\pi) ]$ at large time distances.

Note that when a spatial component of either the pion or the kaon momentum is vanishing, the corresponding matrix element $\langle \pi(p_\pi) | V_i | \pi(p_\pi) \rangle$ or $\langle K(p_K) | V_i | K(p_K) \rangle$ ($i = 1, 2, 3$) is also vanishing and cannot be used in the denominator of the r.h.s~of Eq.~(\ref{eq:RVmu}).
In these cases the quantity $2p_{\pi i} / \langle  \pi(p_\pi) | V_i | \pi(p_\pi) \rangle$ (or $2p_{K i} / \langle  K(p_K) | V_i | K(p_K) \rangle$) is replaced in Eq.~(\ref{eq:RVmu}) by $2E_\pi / \langle  \pi(p_\pi) | V_0| \pi(p_\pi) \rangle$ (or $2E_K / \langle K(p_K) | V_0 | K(p_K) \rangle$).

In the case of both pion and kaon at rest only the time component $R_0^{(V)}$ can be constructed providing a quite precise determination of the scalar form factor $f_0(q^2)$ at kinematical end-point $q^2 = q_{max}^2 \equiv (M_K - M_\pi)^2$, namely \cite{Becirevic:2004ya} 
 \be
     \label{eq:f0qmax}
     R_0^{(V)}(t; \vec{p}_K = \vec{0}, \vec{p}_\pi = \vec{0}) _{ ~ \overrightarrow{t \gg a, ~ (T/2 - t) \gg a} ~ } [(M_K + M_\pi) f_0(q_{max}^2)]^2 ~ .
 \ee
In our present simulations Eq.~(\ref{eq:f0qmax}) allows to achieve a precision for $f_0(q_{max}^2)$ ranging from $\simeq 0.02 \%$ to $\simeq 0.5 \%$ (see later Table \ref{tab:f+0}).

In the case of the scalar density we investigate the time dependence of a suitable double ratio $R^{(S)}$ of 2- and 3-point correlation functions, defined in terms of the local bare density $S(x)$ as
 \be
      \label{eq:RS}
      R^{(S)}(t; \vec{p}_K, \vec{p}_\pi) \equiv 16 E_K E_\pi ~ \frac{C_S^{K \pi}(t, \frac{T}{2}; \vec{p}_K, \vec{p}_\pi) ~ C_S^{\pi K}(t, \frac{T}{2};  \vec{p}_\pi, \vec{p}_K)}
      {C_2^K(T/2; \vec{p}_K) ~ C_2^\pi(T/2; \vec{p}_\pi)}
 \ee
At large time distances $t$ one gets
 \be
    \label{eq:RS_larget}
    R^{(S)}(t; \vec{p}_K, \vec{p}_\pi) _{ ~ \overrightarrow{t \gg a, ~ (T/2 - t) \gg a} ~ } |\langle \pi(p_\pi) | S | K(p_K) \rangle |^2 ~ ,
 \ee
which provides directly the (bare) quantity $|\langle S \rangle|^2$ independently on the matrix elements $Z_\pi$ and $Z_K$ of the interpolating PS fields.
As in the case of the vector current, taking the square root of $R^{(S)}$ the absolute value of the matrix element $\langle S \rangle$ can be obtained, while the sign of $\langle S \rangle$ can be easily inferred from the sign of the plateau of the ratio $C_S^{K \pi}(t, \frac{T}{2}; \vec{p}_K, \vec{p}_\pi) / [ C_2^K(t; \vec{p}_K) ~ C_2^\pi(T/2 - t; \vec{p}_\pi) ]$ at large time distances.

\begin{figure}[htb!]
\begin{center}
\scalebox{0.60}{\includegraphics{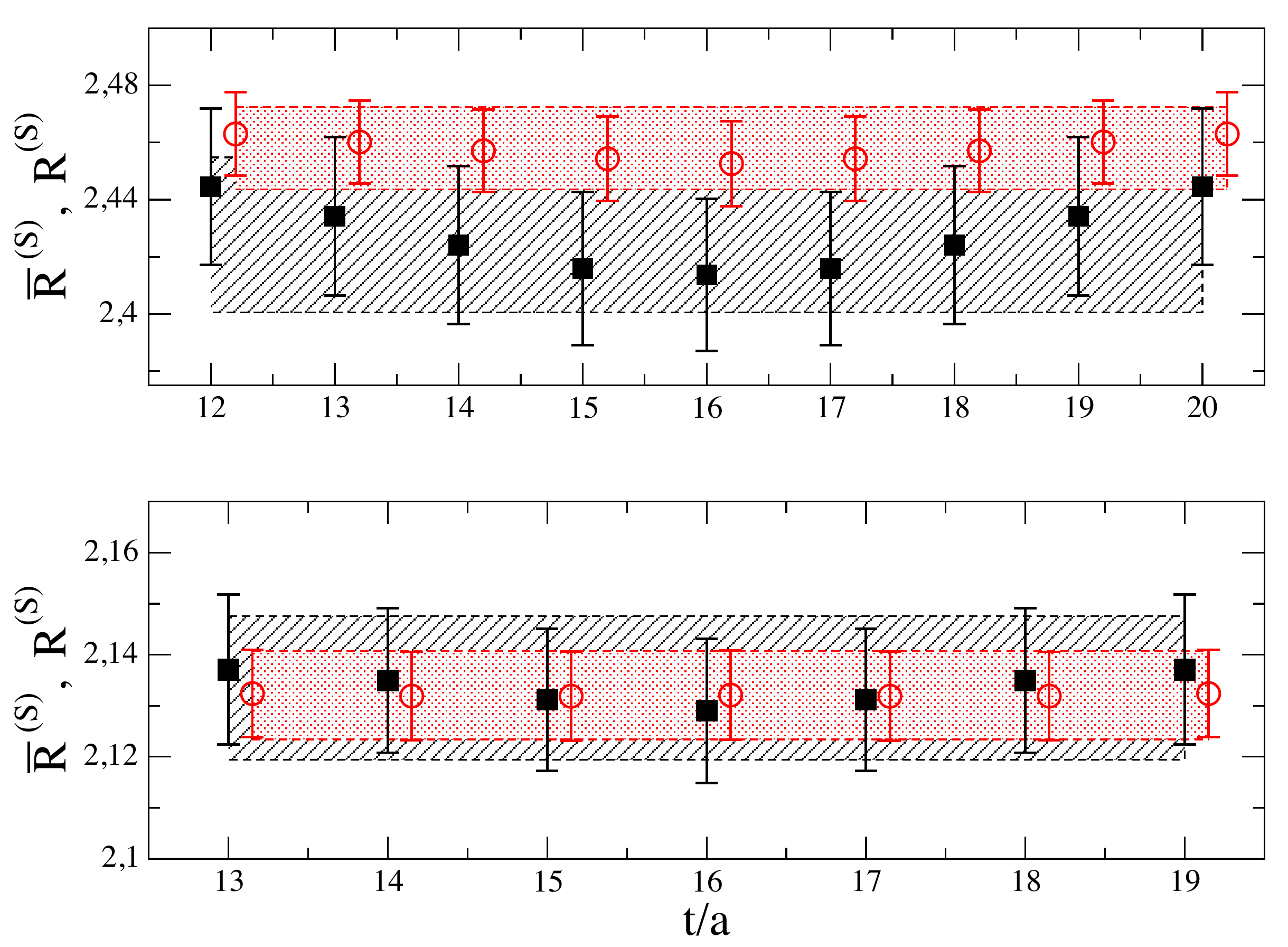}}
\end{center}
\vspace*{-0.75cm}
\caption{\it Double ratios $R^{(S)}$ (black squares) and $\overline{R}^{(S)}$ (red dots), given by Eqs.~(\ref{eq:RS}) and (\ref{eq:RSnew}), respectively, versus the Euclidean time $t / a$ in the cases of the ensembles A30.32 with $a \mu_s = 0.0185$ (upper panel) and B55.32 with $a \mu_s = 0.0180$ (lower panel). The meson masses are $(M_\pi, ~ M_K) \simeq (275, ~ 510)$ MeV (upper panel) and $(M_\pi, ~ M_K) \simeq (375, ~ 545)$ MeV (lower panel). The values of the pion and kaon momentum correspond $|\vec{p}_K| = 0$ and $|\vec{p}_\pi | \simeq 150$ MeV for both panels. The black and red areas correspond to the values of the extracted matrix element $ |\langle \pi(p_\pi) | S | K(p_K) \rangle |^2$ obtained using the plateau regions described later in the text and shown in the figure. The corresponding values of the squared four-momentum transfer $q^2$ are close to $q^2 = 0$, namely $q^2 \simeq 0.016 \gev^2$ (upper panel) and $q^2  \simeq -0.003 \gev^2$ (lower panel).}
\label{fig:sisi}
\end{figure}

When both pion and kaon are at rest Eq.~(\ref{eq:RS}) provides a determination of $f_0(q_{max}^2)$, which turns out to be significantly less precise than the one obtained from Eq.~(\ref{eq:f0qmax}), i.e.~using the vector current.
Moreover, at variance with Eq.~(\ref{eq:RVmu}), where only ratios of 3-point correlation functions appear, the definition (\ref{eq:RS}) involves ratios of 3- to 2-point functions, which 
typically are less correlated each other.
Thus, in order both to make use of 3-point functions only and to take advantage of the statistically precise quantity $R_0^{(V)}(t; \vec{0}, \vec{0})$ we introduce a new {\it scalar} ratio $\overline{R}^{(S)}$, which combines 3-point correlators corresponding to both scalar and vector quark currents, defined as
 \be
     \label{eq:RSnew}
     \overline{R}^{(S)}(t; \vec{p}_K, \vec{p}_\pi) \equiv \overline{K} ~ R_0^{(V)}(t; \vec{0}, \vec{0}) ~ \frac{C_S^{K \pi}(t, \frac{T}{2}; \vec{p}_K, \vec{p}_\pi) ~ 
      C_S^{\pi K}(t, \frac{T}{2};  \vec{p}_\pi, \vec{p}_K)}{C_S^{K \pi}(t, \frac{T}{2}; \vec{0}, \vec{0}) ~ C_S^{\pi K}(t, \frac{T}{2};  \vec{0}, \vec{0})} ~ ,
 \ee
where $\overline{K}$ is a simple kinematical factor depending on meson and quark masses, namely
 \be
     \label{eq:K}
     \overline{K} = \left[ \frac{M_K - M_\pi}{\mu_s - \mu_\ell} \frac{E_K E_\pi}{M_K M_\pi} \right]^2 e^{(E_K - M_K+E_\pi-M_\pi) \dot T / 2} ~ .
 \ee
At large time distances $t$ one gets again (up to lattice artefacts which may differ from those of the ratio $R^{(S)}$) 
 \be
    \label{eq:RSnew_larget}
    \overline{R}^{(S)}(t; \vec{p}_K, \vec{p}_\pi) _{ ~ \overrightarrow{t \gg a, ~ (T/2 - t) \gg a} ~ } | \langle \pi(p_\pi) | S | K(p_K) \rangle |^2 ~ .
 \ee
The use of Eq.~(\ref{eq:RSnew}) improves significantly the statistical precision of the extracted matrix element $\langle S \rangle$ with respect to the case of Eq.~(\ref{eq:RS}), as it is illustrated in Fig.~\ref{fig:sisi}.
The improvement in the precision may reach even a factor larger than $2$.

In lattice QCD simulations the spatial components of the hadronic momenta $p_j$ ($j = 1, 2, 3$) are quantized.
The specific values depend on the choice of the boundary conditions (BC's) applied to the quark fields.
In Refs.~\cite{Bedaque:2004kc,deDivitiis:2004kq} it was proposed to use (partially) twisted BC's for the (valence) quark fields
 \be
    \psi(x + \hat{e}_j L) = e^{2 \pi i \theta_j} ~ \psi(x)
    \label{eq:twistedBC}
 \ee
which allow to shift the quantized values of $p_j$ by an arbitrary amount equal to $2 \pi \theta_j / L$, namely
 \be
    p_j = \theta_j \frac{2 \pi}{L} + n_j \frac{2 \pi}{L} ~ ,
    \label{eq:pj_twisted}
  \ee
where the $n_j$'s are integer numbers.
Such a choice is crucial for removing the strong limitations to the kinematical regions accessible for the investigation of momentum dependent quantities (like e.g.~form factors), imposed by the use of periodic BC's (i.e.~$\theta_j = 0$).
In Refs.~\cite{Guadagnoli:2005be,Flynn:2005in} it was shown that the momentum shift produced by the partially twisted BC's does not introduce any additional noise and allows to determine the form factors of the $K_{\ell 3}$ decays with good accuracy for both spacelike and timelike $4-$momentum transfer $q^2$.

The partially twisted BC's can be tuned to determine the $K_{\ell 3}$ form factors directly at the relevant kinematical point $q^2 = 0$ \cite{Boyle:2007wg,Boyle:2010bh}, avoiding in this way the need for an Ansatz with which to perform the $q^2$-interpolation of the form factor data\footnote{The systematic uncertainty related to the choice of the Ansatz for the $q^2$-interpolation turns out however to be a sub-dominant effect (see Ref.~\cite{Lubicz:2009ht} and later on Section \ref{sec:first_strategy}).}.
Such a direct strategy has been used in the recent computations of $f_+(0)$ performed by the FNAL/MILC \cite{Bazavov:2013maa} and RBC/UKQCD \cite{Boyle:2015hfa} collaborations.
In this work, instead, we study the momentum dependence of both the scalar and vector form factors in order to compare it with the experimental data, while profiting at the same time of the high precision achievable for the scalar form factor near the kinematical end-point $q^2 = q_{max}^2$ (see Refs.~\cite{Becirevic:2004ya,Guadagnoli:2005be,Lubicz:2009ht}).
Note in particular that, thanks to the use of the new ratio (\ref{eq:RSnew}), the high-precision determination of $f_0(q_{max}^2)$ from the vector quark current improves significantly the evaluation of the scalar form factor $f_0(q^2)$ from the scalar density also at $q^2 \simeq 0$ (see Fig.~\ref{fig:sisi}). 

According to Ref.~\cite{Frezzotti:2003ni} the matrix elements $\langle \hat{V}_\mu \rangle$ and $\langle S \rangle$ can be ${\cal{O}}(a)$-improved by averaging the two kinematics with opposite spatial momenta of the initial and final mesons, namely
 \bea
       \label{eq:improvement_V0}
       \langle \hat{V}_0 \rangle_{imp} & \equiv & \frac{1}{2} \left[ \langle \pi(E_\pi, \vec{p}_\pi) | \hat{V}_0 | K(E_K, \vec{p}_K) \rangle + 
       \langle \pi(E_\pi, - \vec{p}_\pi) | \hat{V}_0 | K(E_K, -\vec{p}_K) \right] ~ , \\
       \label{eq:improvement_Vi}
       \langle \hat{V}_i \rangle_{imp} & \equiv & \frac{1}{2} \left[ \langle \pi(E_\pi, \vec{p}_\pi) | \hat{V}_i | K(E_K, \vec{p}_K) \rangle - 
       \langle \pi(E_\pi, - \vec{p}_\pi) | \hat{V}_i | K(E_K, -\vec{p}_K) \right] ~ , \\
       \label{eq:improvement_S}
       \langle S \rangle_{imp} & \equiv & \frac{1}{2} \left[ \langle \pi(E_\pi, \vec{p}_\pi) | S | K(E_K, \vec{p}_K) \rangle + 
       \langle \pi(E_\pi, - \vec{p}_\pi) | S  | K(E_K, -\vec{p}_K) \right] ~ . 
 \eea
The effect of the above averages turns out to be quite mild (within the statistical errors) in the case of both the time component of the vector current and the scalar density, while it is found to be significant for the matrix elements of the spatial components of the vector current.

We impose on the valence quark fields twisted BC's in the spatial directions and anti-periodic BC's in time.
On the other hand the ETMC gauge configurations have been produced by including in the sea dynamical quarks with periodic BC's in space and anti-periodic ones in time. 
It has been shown \cite{Sachrajda:2004mi,Bedaque:2004ax} that for physical quantities, which do not involve final state interactions (like, e.g., meson masses, decay constants, semileptonic form factors and e.m.~transitions), the use of different BC's for valence and sea quarks is legitimate, since it produces finite volume effects which are exponentially small.

Thus, generalizing Eq.~(\ref{eq:twistedBC}) to four dimensions, we introduce the four-vector $\theta_\mu$ given by
 \be
      \label{eq:theta}
      \theta_\mu = \left( \frac{L}{T} \theta_0, \theta_1, \theta_2, \theta_3 \right) = \left( \frac{L}{2T}, \theta, \theta, \theta \right) ~ ,
 \ee
where the values of $\theta$ adopted for the various gauge ensembles are collected in Table \ref{tab:theta}. 
They have been chosen in order to provide approximately the same physical values of the quark spatial momenta $\vec{p} = (2 \pi / L) \left( \theta, \theta, \theta \right)$ at the various lattice spacings and volumes.
In addition, the opposite values of $\theta$ allow to perform the averages of Eqs.~(\ref{eq:improvement_V0}-\ref{eq:improvement_S}) achieving in this way the ${\cal{O}}(a)$-improvement for the matrix elements.

\begin{table}[htb!]
\begin{center}
\begin{tabular}{||c|c||c||}
\hline
$\beta$ & $V/a^4$ & $\theta$  \\
\hline
~ $1.90$ ~ & ~ $24^3 \times 48$ ~ & $~ \{-0.350, -0.150, 0.0, 0.150, 0.350\}$ ~ \\
$1.90$ & $32^3 \times 64$ & $\{-0.467, -0.200, 0.0, 0.200, 0.467\}$ \\ \hline
$1.95$ & $24^3 \times 48$ & $\{-0.321, -0.138, 0.0, 0.138, 0.321\}$ \\
$1.95$ & $32^3 \times 64$ & $\{-0.427, -0.183, 0.0, 0.183, 0.427\}$ \\ \hline
$2.10$ & $48^3 \times 96$ & $\{-0.493, -0.212, 0.0, 0.212, 0.493\}$  \\
\hline
\end{tabular}
\end{center}
\caption{\it Values of the parameter $\theta$ appearing in Eq.~(\ref{eq:theta}) chosen for the various ETMC gauge ensembles with $N_f = 2+1+1$ adopted in this work.} 
\label{tab:theta}
\end{table}

For the calculation of the relevant 3-point correlation functions $C_{V_\mu}^{K \pi}(t, \frac{T}{2}; \vec{p}_K, \vec{p}_\pi)$ and $C_S^{K \pi}(t, \frac{T}{2}; \vec{p}_K, \vec{p}_\pi)$ we consider the spectator $d$-quark at rest and apply the partially twisted BC's (\ref{eq:theta}) to the initial $s$- and final $u$-quarks.
In this way the kaon and pion momentum are given by $\vec{p}_K = (2 \pi / L) \left( \theta_K, \theta_K, \theta_K \right)$ and $\vec{p}_\pi = (2 \pi / L) \left( \theta_\pi, \theta_\pi, \theta_\pi \right)$, where $\theta_K$ and $\theta_\pi$ assume the values of the parameter $\theta$ given in Table \ref{tab:theta} for each gauge ensemble. 

Since we are using spatially symmetric twisted BC's, the matrix elements of the spatial components of the vector current $\langle \hat{V}_i \rangle_{imp}$ are equal to each other. 
Therefore, in order to improve the statistics, we average them to get
 \be
       \label{eq:Vsp}
       \langle \hat{V}_{sp} \rangle_{imp} = \frac{1}{3} \left[  \langle \hat{V}_1 \rangle_{imp} +  \langle \hat{V}_2 \rangle_{imp} +  \langle \hat{V}_3 \rangle_{imp} \right]
 \ee
 
\begin{figure}[htb!]
\begin{center}
\scalebox{0.60}{\includegraphics{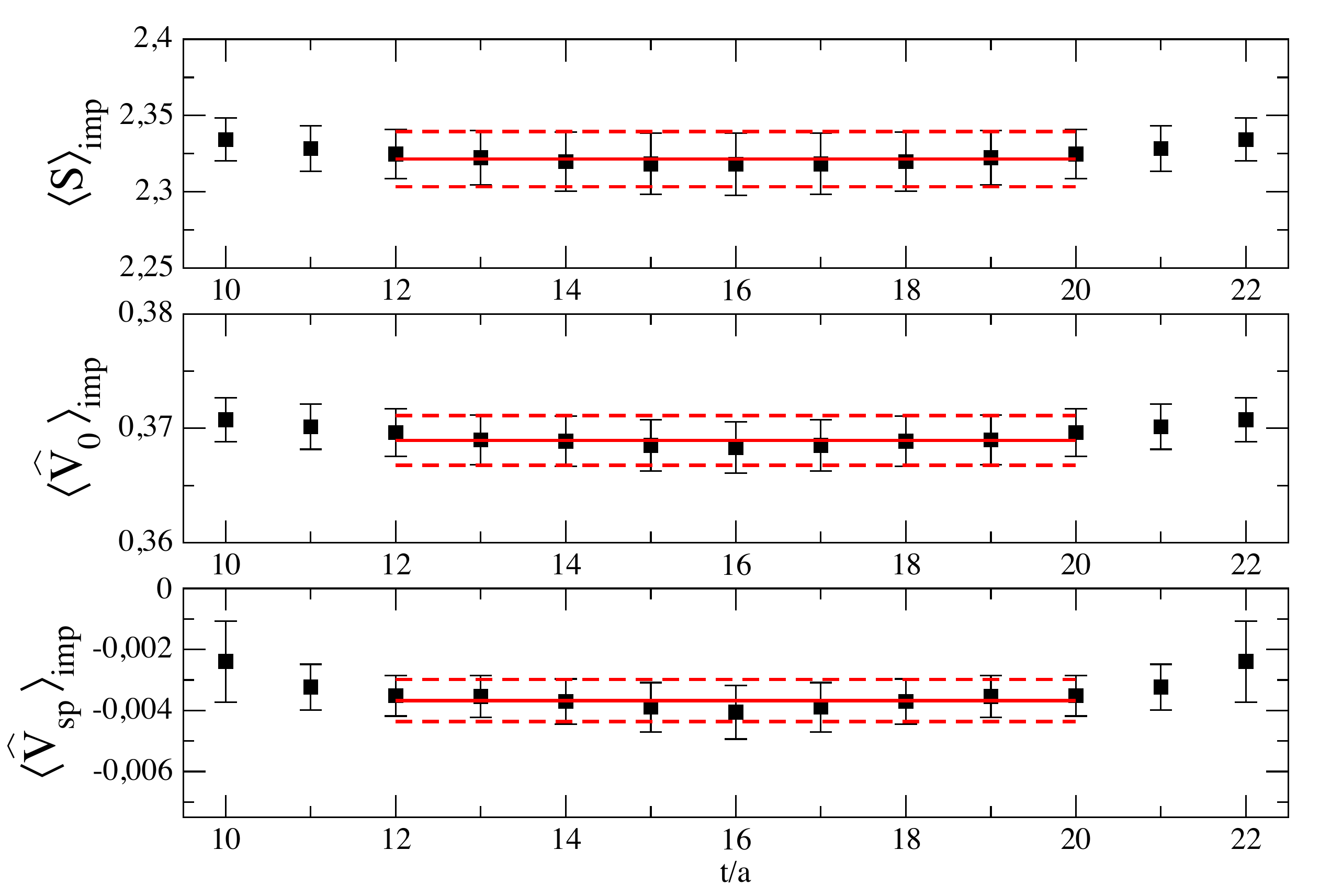}}
\end{center}
\vspace*{-0.75cm}
\caption{\it Matrix elements $\langle \hat{V}_{sp} \rangle_{imp}$, $\langle \hat{V}_{0} \rangle_{imp}$ and $\langle S \rangle_{imp}$ extracted from the ratios $R_{\mu}$ and $\overline{R}_S$ [see Eqs.~(\ref{eq:RVmu}) and (\ref{eq:RSnew})] for the ensemble A40.32 with $\beta = 1.90$, $L / a = 32$, $a\mu_l = 0.0040$, $a\mu_s = 0.0185$, $\vec{p}_K = - \vec{p}_\pi$ and $|\vec{p}_K| \simeq 150 \mev$. The meson masses are $M_\pi \simeq 315$ MeV and $M_K \simeq 525$ MeV. The horizontal red lines correspond to the plateau regions used to extract the matrix elements and to their central values and statistical errors (see text).}
\label{fig:matel}
\end{figure}

The nice quality of the plateaux for the matrix elements $\langle \hat{V}_0 \rangle_{imp}$, $\langle \hat{V}_{sp} \rangle_{imp}$ and $\langle S \rangle_{imp}$ is illustrated in Fig.~\ref{fig:matel}.
The time intervals adopted for the fits (\ref{eq:RVmu_larget}) and (\ref{eq:RSnew_larget}) are symmetric around $T/4a$ and equal to ($10 - 14$) for $T/a = 48$, ($12 - 20$) for $T/a = 64$ at $\beta = 1.90$, ($13 - 19$) for $T/a = 64$ at $\beta = 1.95$ and ($18 - 30$) for $T/a = 96$.
These values are compatible with the dominance of the pion and kaon ground-state observed in the two-point correlation functions in Ref.~\cite{Carrasco:2014cwa}.
We have also checked that the more conservative choice $[T/4a - 2, T/4a + 2]$ for the lattices with time extension $T/a = 64$ and $96$ yield changes in the extracted matrix elements well below the statistical uncertainty\footnote{Excited states may contaminate the double ratios (\ref{eq:RVmu_larget}) and (\ref{eq:RSnew_larget}) in a $t$-independent way. However in the case of the ratio $C_{\hat{\Gamma}}^{K \pi}(t, t^\prime; \vec{p}_K, \vec{p}_\pi) / C_2^K(t; \vec{p}_K) C_2^\pi(t' - t; \vec{p}_\pi)$ the contaminations are $t$-dependent, since the excited staes from the source and the sink are different. We have checked that for the above ratio the quality of the plateaux is similar to the one shown in Figs.~\ref{fig:sisi}-\ref{fig:matel}.}.

To sum up, from the 2-point and 3-point lattice correlators we determine three ${\cal{O}}(a)$-improved matrix elements, namely $\langle \hat{V}_0 \rangle_{imp}$, $\langle \hat{V}_{sp} \rangle_{imp}$ and $\langle S \rangle_{imp}$, which are related to the scalar $f_0(q^2)$ and vector $f_+(q^2)$ form factors by the redundant system of relations given by
 \bea
       \label{eq:V0_final}
       \langle \hat{V}_0 \rangle_{imp} & = & (E_K + E_\pi) f_+(q^2) + (E_K - E_\pi) \frac{M_K^2 - M_\pi^2}{q^2} \left[ f_0(q^2) - f_+(q^2) \right] + {\cal{O}}(a^2)~ , \\[2mm]
       \label{eq:Vsp_final}
       \langle \hat{V}_{sp} \rangle_{imp} & = & \frac{2 \pi}{L} \left\{ (\theta_K + \theta_\pi) f_+(q^2) + (\theta_K - \theta_\pi) \frac{M_K^2 - M_\pi^2}{q^2} 
                                                                       \left[ f_0(q^2) - f_+(q^2) \right] \right\} + {\cal{O}}(a^2) ~ , \qquad \\[2mm]
       \label{eq:S_final}
       \langle S \rangle_{imp}  & = & \frac{M_K^2 - M_\pi^2}{\mu_s - \mu_\ell} f_0(q^2) + {\cal{O}}(a^2) ~ .
 \eea
We determine the form factors $f_0(q^2)$ and $f_+(q^2)$ by minimizing the $\chi^2$-variable constructed using the three Eqs.~(\ref{eq:V0_final}-\ref{eq:S_final}).

After a small interpolation of our lattice data to the physical value of the strange quark mass $m_s = 99.6 (4.3) \mev$, determined in Ref.~\cite{Carrasco:2014cwa}, we present in the next Sections our results at the physical point and in the continuum limit coming from two different strategies.

The first strategy is based on the study of the momentum dependence of the form factors $f_0(q^2)$ and $f_+(q^2)$ using either the $z-$expansion of Ref.~\cite{Bourrely:2008za} or a polynomial fit.
In this way we determine the form factor $f_+(0)$ at each simulated value of the (renormalized) light-quark mass $m_\ell$.
Then the results obtained for $f_+(0)$ are extrapolated to the physical point $m_\ell = m_{ud}$ and to the continuum limit using either SU(2) \cite{Flynn:2008tg} or SU(3) \cite{Gasser:1984ux,Gasser:1984gg} Chiral Perturbation Theory (ChPT) predictions.

The second strategy, inspired by our previous work \cite{Lubicz:2010bv} done with the $N_f = 2$ ETMC gauge ensembles, is based on a combined fit of the $q^2$, $m_{\ell}$ and lattice spacing dependencies of our lattice data for the form factors $f_0(q^2)$ and $f_+(q^2)$ together with the result for the ratio of the kaon and pion leptonic decay constants $f_K / f_\pi$ obtained at the chiral point in Ref.~\cite{Carrasco:2014poa}.
The latter quantity is used for imposing the constraint coming from the Callan-Treiman theorem \cite{Callan:1966hu}.
In this way the momentum dependence of the form factors $f_0(q^2)$ and $f_+(q^2)$ is obtained, at the physical point, in the whole range of values of $q^2$ measured in $K_{\ell 3}$ decays~\cite{FlaviaNet,Moulson:2014cra}, i.e.~from $q^2 = 0$ up to the physical end-point $q^2 = q_{max}^2 \simeq 0.129 \gev^2$.

\section{First strategy: interpolation of the form factors at $q^2 = 0$}
\label{sec:first_strategy}

In the first strategy, for each gauge ensemble we fit simultaneously our lattice data for $f_{+,0}(q^2)$ using the $z$-expansion as parametrized in Ref.~\cite{Bourrely:2008za}, namely
 \be
        \label{eq:zexpansion}
         f_{+,0}(q^2 ) = \frac{a_{+,0}^{(0)}  + a_{+,0}^{(1)} \left( z + \frac{1}{2} z^2 \right)}{1 - \frac{q^2}{M_{V,S}^2}} ~ ,
 \ee
where $a_{+,0}^{(0)}$, $a_{+,0}^{(1)}$, $M_V$ and $M_S$ (representing the vector and scalar pole masses, respectively,) are kept as free parameters and $z$ is defined as 
 \be
    z = \frac{{\sqrt {t_ +   - q^2 }  - \sqrt {t_ +   - t_0 } }}{{\sqrt {t_ +   - q^2 }  + \sqrt {t_ +   - t_0 } }}
 \ee
with $t_+$ and $t_0$ given by
 \bea
        t_+  & = & \left( M_K  + M_\pi \right)^2 ~ , \nonumber \\
        t_0  & = & \left( M_K  + M_\pi \right) \left( \sqrt {M_K}  - \sqrt {M_\pi} \right)^2 ~ .
\end{eqnarray}
The condition $f_+(0) = f_0(0)$ is imposed by rewriting Eq.~(\ref{eq:zexpansion}) in the form
 \be
        \label{eq:zexpansion_ff}
         f_{+,0}(q^2 ) = \frac{f_+(0)  + a_{+,0}^{(1)} ~ (z - z_0) \left[ 1 + (z + z_0) / 2 \right]}{1 - q^2 / M_{V,S}^2} ~ ,
 \ee
 where $z_0 \equiv z(q^2 = 0)$, reducing in this way the number of free parameters to five for each gauge ensemble.

\begin{figure}[htb!]
\centering
\scalebox{0.32}{\includegraphics{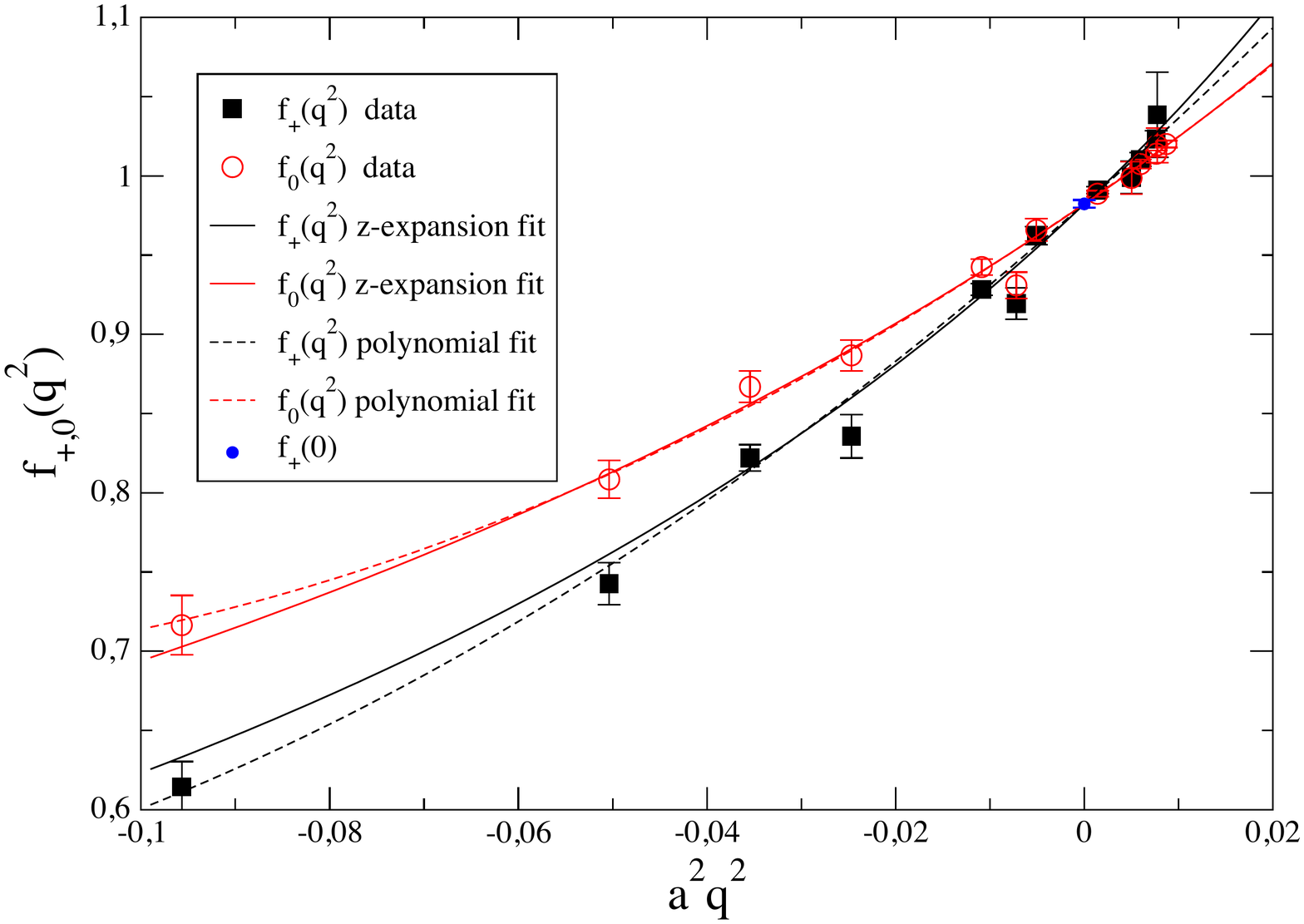}}
\scalebox{0.32}{\includegraphics{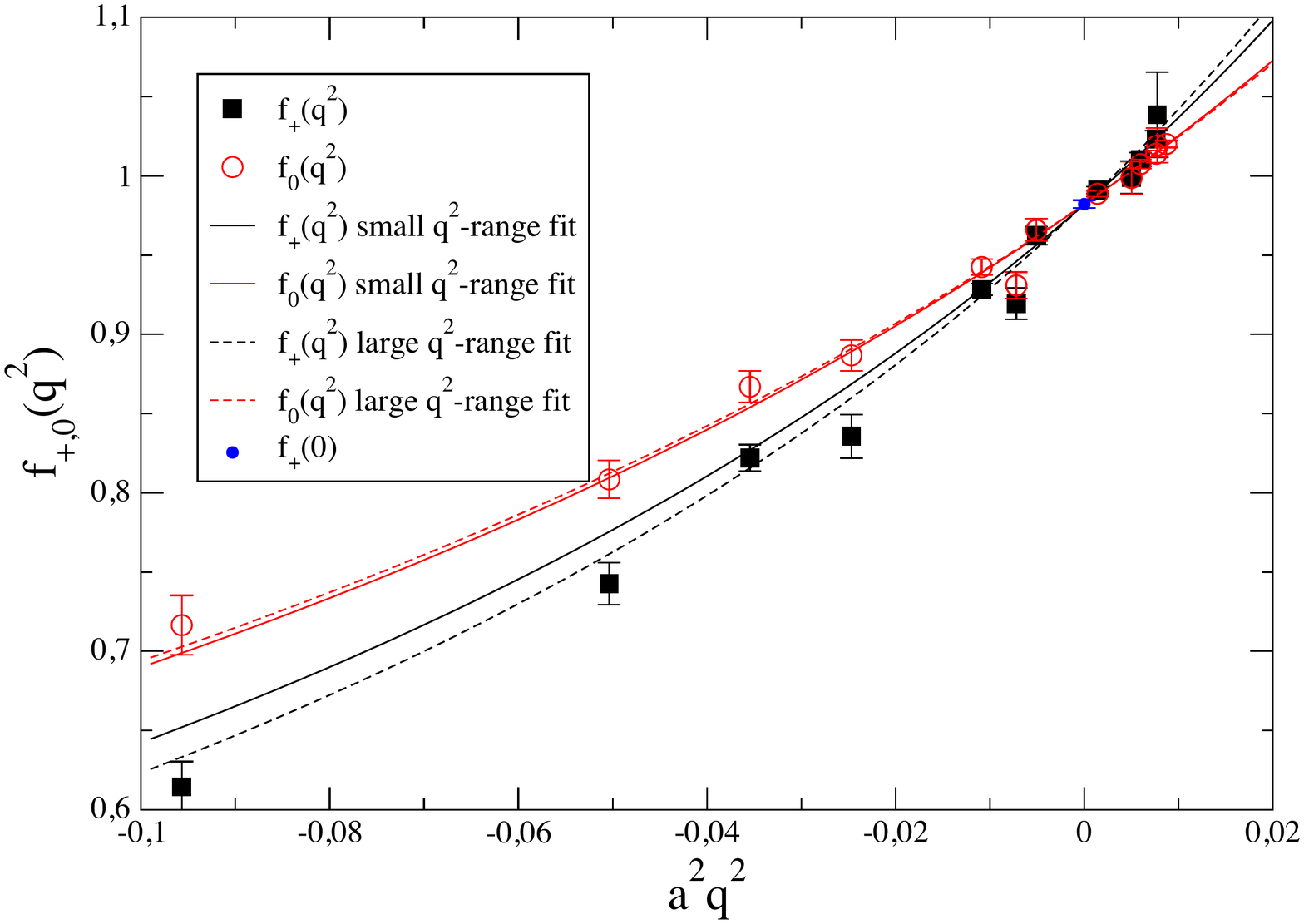}}
\vspace*{-1cm}
\caption{\it Left panel: results of the fit of our lattice data for the form factors $f_+(q^2)$ and $f_0(q^2)$ using the $z$-expansion (continuum lines), given by Eq.~(\ref{eq:zexpansion_ff}), compared to the ones obtained adopting a simple quadratic formula in $q^2$ (dashed lines). Right panel: $z$-expansion fit of the form factors using only the data corresponding to $a^2 q^2 > -0.01$ (continuum lines), or using all the data (dashed lines). Both panels correspond to the ensemble A60.24 with $\beta=1.90$, $L/a = 24$, $a\mu_\ell = 0.0060$ and $a \mu_s = 0.0185$.}
\label{fig:q2fit}
\end{figure}

We also fit the $q^2$-dependence of the form factors using an alternative, simple Ansatz (a quadratic fit in $q^2$), obtaining nearly identical results, as it can be seen in the left panel of Fig.~\ref{fig:q2fit}.
The $q^2$-range of our lattice data include the timelike region $0 < q^2 < q_{max}^2$ as well as the spacelike one, reaching quite large negative values of $q^2$.
Since concerns may be raised about the size of lattice artefacts at large negative values of $q^2$, which correspond to large values of the quark momenta \cite{Boyle:2007wg}, we have either included or excluded in our fits the data corresponding to large negative values of $q^2$ (corresponding to $a^2 q^2 < -0.01$).
Again nearly identical results are obtained, as it is shown in the right panel of Fig.~\ref{fig:q2fit}.

The results for $f_+(0)$ obtained using the $z$-expansion or the quadratic fit in $q^2$ are collected in Table \ref{tab:f+0} for each value of the (renormalized) light-quark mass.
It can clearly be seen that the values of $f_+(0)$ corresponding to the two different ans\"atze for the $q^2$-dependence differ by less than $1/4$ of the statistical errors, which represents therefore a sub-dominant effect (see also Ref.~\cite{Lubicz:2009ht}).
Notice also that: ~ i) the two lattice points calculated at the same lattice spacing and light-quark mass but different volumes (corresponding to the ensembles A40.24 and A40.32) are compatible within $1$ standard deviation, and ~ ii) finite size effects (FSEs) are expected to be maximal for the ensemble A40.24, having the coarsest lattice spacing with the smallest pion mass and lattice volume (see Ref.~\cite{Carrasco:2014cwa}). 

\begin{table}[htb!]
\begin{center}
\begin{tabular}{||c|c|c|c||c|c|c|c||}
\hline
ensemble & $\beta$ & $V / a^4$ &$a\mu_{sea}=a\mu_\ell$&$m_\ell ~ (\mev)$&$f_0(q_{max}^2)$&$f_+(0)$& $f_+(0)$ \\
 & & & & & & (z-expansion) & (quadratic) \\
\hline \hline
$A30.32$ & $1.90$ & $32^3\times 64$ &$0.0030$ & $12.1$ & $1.0397 (19)$ & $0.9792 (28)$ & $0.9804 (27)$ \\
$A40.32$ & & & $0.0040$ & $16.2$ & $1.0250 (28)$ & $0.9767 (32)$ & $0.9774 (31)$ \\
$A50.32$ & & & $0.0050$ & $20.2$ & $1.0109 (13)$ & $0.9813 (15)$ & $0.9808 (15)$  \\
\hline 
$A40.24$ & $1.90$ & $24^3\times 48 $ & $0.0040$ & $16.2$ & $1.0269 (32)$ & $0.9801 (28)$& $0.9802 (26)$ \\
$A60.24$ & & & $0.0060$ & $24.2$ & $1.0101 (12)$ & $0.9870 (16)$ & $0.9865 (16)$ \\
$A80.24$ & & & $0.0080$ & $32.3$ & $1.0025 (04)$ & $0.9924 (04)$ & $0.9922 (04)$ \\
$A100.24$ & & & $0.0100$ & $40.4$ & $1.0005 (03)$ & $0.9969 (04)$ & $0.9969 (03)$ \\
\hline
$B25.32$ & $1.95$ & $32^3\times 64$ &$0.0025$& $11.5$ & $1.0486 (35)$ & $0.9772 (50)$ & $0.9767 (51)$ \\
$B35.32$ & & & $0.0035$ & $16.1$ & $1.0222 (20)$ & $0.9808 (28)$ & $0.9810 (28)$ \\
$B55.32$ & & & $0.0055$ & $25.3$ & $1.0095 (07)$ & $0.9910 (12)$ & $0.9906 (13)$ \\
$B75.32$ & & & $0.0075$ & $34.5$ & $1.0028 (04)$ & $0.9938 (10)$ & $0.9938 (10)$ \\
\hline
$B85.24$ & $1.95$ & $24^3\times 48 $ & $0.0085$ & $39.1$ & $1.0012 ~(2)$ & $0.9963 (02)$ & $0.9964 (03)$  \\
\hline
$D15.48$ & $2.10$ & $48^3\times 96$ &$0.0015$& $~9.0$ & $1.0567 (55)$ & $0.9833 (84)$& $0.9841 (83)$ \\ 
$D20.48$ & & & $0.0020$ & $12.0$ & $1.0400 (32)$ & $0.9842 (42)$ & $0.9844 (46)$ \\
$D30.48$ & & & $0.0030$ & $18.1$ & $1.0162 (16)$ & $0.9870 (24)$ & $0.9863 (26)$ \\
 \hline   
\end{tabular}
\end{center}
\caption{\it Values of the scalar form factor $f_0(q_{max}^2)$ at the kinematical end-point $q_{max}^2 = (M_K - M_\pi)^2$ and of the form factor $f_+(0) = f_0(0)$, obtained using the $z$-expansion (\ref{eq:zexpansion_ff}) or a simple quadratic fit in $q^2$, versus the (renormalized) light-quark mass $m_\ell$ for each gauge ensemble used in this work. The (renomalized) strange quark mass has been interpolated at the physical value $m_s = 99.6 (4.3) \mev$~\cite{Carrasco:2014cwa}.}
\label{tab:f+0}
\end{table}

In order to compute the physical value of the form factor $f_+(0)$, we perform the extrapolation to the physical point using both SU(2) \cite{Flynn:2008tg} and SU(3) ChPT predictions \cite{Gasser:1984ux,Gasser:1984gg}.

For the SU(2) ChPT Ansatz we use
 \be
    \label{eq:SU2fit}
    f_ +  (0) = F_+ \left[ 1 - \frac{3}{4}\xi \log \xi  + C_1 \xi + C_2 \xi^2  + D (a / r_0)^2 \right] ~ ,
 \ee
where $\xi_\ell \equiv 2B m_\ell / 16\pi^2f^2$ with $B$ and $f$ being the SU(2) low-energy constants (LECs) entering the LO chiral Lagrangian and determined in Ref.~\cite{Carrasco:2014cwa}, while the quantities $F_+$, $C_1$, $C_2$ and $D$ are free fitting parameters.
In Eq.~(\ref{eq:SU2fit}) we have added quadratic terms in $\xi$, $C_2 \xi^2$, and in the lattice spacing, $D_+ (a / r_0)^2$, in order to take into account  next-to-next-to-leading order (NNLO) ChPT and ${\cal{O}}(a)$-improved discretization effects, respectively.

The SU(3) ChPT expansion of the form factor $f_+(0)$ reads as 
 \be
    \label{eq:SU3fit}
    f_ +  (0) = 1 + f_2  + \Delta f ~ ,  
 \end{equation}
where the NLO term $f_2$ does not depend on any NLO LEC and it can be written in terms of meson masses, namely \cite{Gasser:1984ux,Gasser:1984gg}
 \be
     \label{eq:f2}
     f_2 = \frac{3}{2}H_{\pi K}  + \frac{3}{2}H_{\eta K} ~ ,  
 \ee
with
 \be
     \label{eq:HPQ}
     H_{PQ}  =  - \frac{1}{64 \pi^2 f^2} \left[ M_P^2  + M_Q^2  + \frac{2M_P^2 M_Q^2}{M_P^2  - M_Q^2} \log \frac{M_Q^2}{M_P^2} \right] ~ . 
 \ee
In Eq.~(\ref{eq:SU3fit}) the quantity $\Delta f$ represents NNLO contributions and beyond, which we parametrize in our fit as 
 \be
    \label{eq:Deltaf}
    \Delta f = \left( \frac{M_K^2 - M_\pi^2}{M_K^2} \right)^2 \left[ \Delta_0  + \Delta_1 \xi+ \Delta_2 (a / r_0)^2 \right] ~ ,
 \ee
where $\Delta_{0, 1, 2}$ are free parameters.
Equations~(\ref{eq:SU3fit}-\ref{eq:Deltaf}) verify the constraint imposed by the Ademollo-Gatto theorem \cite{Ademollo:1964sr} at any value of the lattice spacing, i.e.~deviations of $f_+(0)$ from unity are at least quadratic in the SU(3)-breaking parameter $(M_K^2 - M_\pi^2) \propto (m_s - m_\ell)$.
We tried also an alternative Ansatz for $\Delta f$ in which the Ademollo-Gatto theorem is satisfied only in the continuum limit, namely $\Delta f =  \left[ \tilde{\Delta}_0  + \tilde{\Delta}_1 \xi \right] \left( M_K^2 - M_\pi^2 \right)^2 / M_K^4+ \tilde{\Delta}_2 (a / r_0)^2$, obtaining however a fit of poorer quality to the lattice data.

Notice that: ~ i) in the SU(3) fit of our lattice data the strange quark mass is fixed at its physical value, and ~ ii) in Eq.~(\ref{eq:HPQ}) the pion decay constant $f$ at the chiral point can be replaced by its value $f_\pi$ at the physical point. 
The difference is a NNLO effect that should be reabsorbed by $\Delta f$. 
We have verified that the final result for $f_+(0)$ at the physical point is almost independent on the precise choice of the pion decay constant $f$ appearing in Eq.~(\ref{eq:HPQ}), once Eq.~(\ref{eq:Deltaf}) is included in the fit.

\begin{figure}[htb!]
\centering
\scalebox{0.32}{\includegraphics{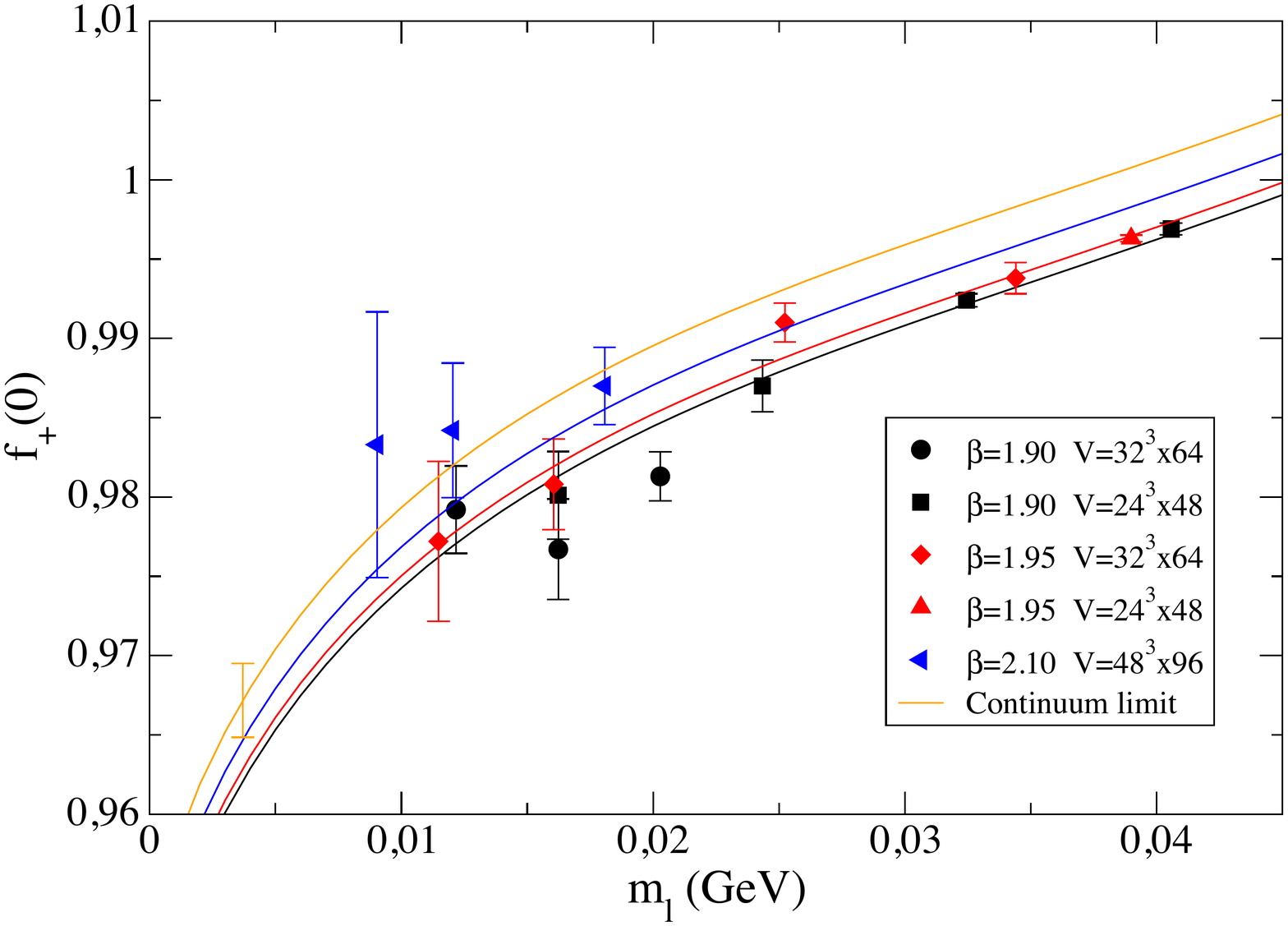}}
\scalebox{0.32}{\includegraphics{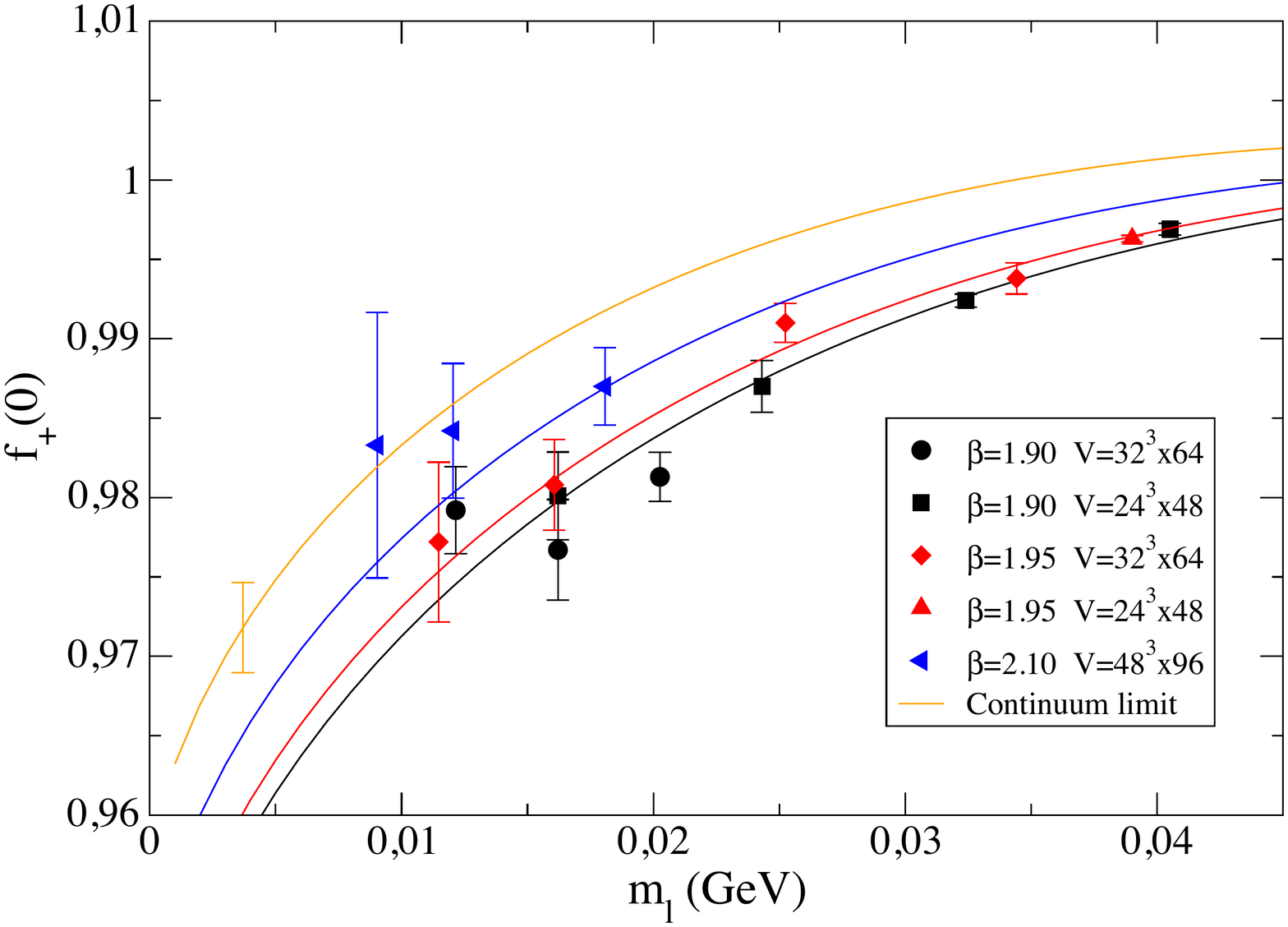}}
\vspace*{-0.5cm}
\caption{\it Chiral and continuum extrapolation of $f_+(0)$ based on the SU(2) ChPT fit given by Eq.~(\ref{eq:SU2fit}) (left) and on the SU(3) ChPT fit given by Eqs.~(\ref{eq:SU3fit}-\ref{eq:Deltaf}) (right).}
\label{fig:SU2SU3fit}
\end{figure}

The chiral and continuum extrapolations of $f_+(0)$ are shown in Fig.~\ref{fig:SU2SU3fit} for both the SU(2) and SU(3) fits. 
Combining the two analysis we get our result for the vector form factor $f_+(0)$ at the physical point and in the continuum limit:
 \be
    \label{eq:f+0_final1}
    f_+(0) = 0.9697 ~ (45)_{stat+fit} ~ (11)_{input} ~ (25)_{syst} = 0.9697 ~ (52) ~ ,
 \ee
where $()_{stat+fit}$ indicates the statistical uncertainty which includes also the one induced by the fitting procedure, while $()_{input}$ takes into account the uncertainties present in the determination of the input parameters \cite{Carrasco:2014cwa}, namely the values of the average $u/d$ quark mass $m_{ud}$, the lattice spacing $a$ and the SU(2) ChPT low energy constants $f$ and $B$.
The systematic uncertainty, indicated as $()_{syst}$, is dominated by the chiral extrapolation error, estimated as half of the difference of the results corresponding to the SU(2) and SU(3) ChPT extrapolations.
The discretization error has been estimated by comparing with the results obtained including in Eqs.~(\ref{eq:SU2fit}) and (\ref{eq:Deltaf}) a term proportional to $(a/r_0)^4$ (adopting for the coefficient a prior equal to $0 \pm 3$), and turns out to be below the permille level. 

Before closing this section we note that for each bootstrap event we use the uncorrelated $\chi^2$-variable for our fits. 
The $\chi^2$ is not used to estimate the errors on the fitting parameters, which are always evaluated with the boostrap procedure, in which correlations are automatically taken into account. 
Since part of the uncertainties in our fits have also a systematic origin, like the distribution of the input parameters in the analyses carried out in Ref.~\cite{Carrasco:2014cwa}, the $\chi^2$-distribution is not fully significant in order to quantify the quality of the extrapolations.
Indeed, the goodness-of-fit tests would require statistical errors only.
Nevertheless, we note that the values of $\chi^2$ in the various fits carried out for the chiral and continuum extrapolations turn out to comparable. 
This is the reason why we combine the results corresponding to the various branches of the analysis with the same weight, according to Eq. (28) of Ref.~\cite{Carrasco:2014cwa}.

\section{Second strategy: combined fit}
\label{sec:second_strategy}

In this Section we present the results of our second strategy in which we extrapolate the form factors $f_+(q^2)$ and $f_0(q^2)$ to the physical point for a wide range of values of $q^2$, which includes the $q^2$-region accessible to experiments, i.e.~from $q^2 = 0$ to the kinematical end-point $q^2 = q_{max}^2  \simeq 0.129 \gev^2$. 

We perform a combined fit of the $q^2$-, $m_\ell$- and $a$-dependencies of the form factors using either the SU(2) ChPT inspired prediction proposed in Ref.~\cite{Lubicz:2010bv} or a modified $z$-expansion.
In both cases we include the constraint arising from the CT theorem \cite{Callan:1966hu}, which relates the scalar form factor $f_0(q^2)$ calculated at the unphysical point $q^2 = q_{CT}^2 = M_K^2 - M_\pi^2$ to the ratio of the leptonic decay constants $f_K / f_\pi$ in the SU(2) chiral limit.

\subsection{SU(2) ChPT inspired Ansatz}
\label{sec:SU2}

Following Ref.~\cite{Lubicz:2010bv} we derive the first Ansatz based on SU(2) ChPT starting from the expansion of the NLO SU(3) ChPT predictions for the semileptonic form factors \cite{Gasser:1984ux,Gasser:1984gg} in powers of the variable $x \equiv M_\pi^2 / M_K^2$, keeping only the ${\cal{O}}(x)$, ${\cal{O}}(x \log x)$ and ${\cal{O}}(\log(1 - s))$ terms, where $s \equiv q^2 / M_K^2$.
The result of the expansion can be cast in the following form
 \be
   \label{eq:ff_SU3exp}
   f_{+,0}(q^2) = F_+ \left\{ 1 + C_+ x - \frac{M_K^2}{(4\pi f)^2} \left[ \frac{3}{4} x \log{x} + x T_{+,0}^{(1)}(s) + T_{+,0}^{(2)}(s) \right] \right\} ~ + ~ \cdots
 \ee       
with 
 \bea
   \label{eq:T+0}
   T_+^{(1)}(s) & = & \left[ (1 - s) \log{(1 - s)} + s(1 - s/2) \right] 3(1 + s) / 4s^2 ~ , \nonumber \\ 
   T_+^{(2)}(s) & = &  \left[ (1 - s) \log{(1 - s)} + s(1 - s/2) \right] (1 - s)^2 / 4s^2 ~ , \nonumber \\
   T_0^{(1)}(s) & = & \left[  \log{(1 - s)} + s(1 + s/2) \right] (9 + 7s^2) / 4s^2 ~ , \nonumber \\
   T_0^{(2)}(s) & = & \left[ (1 - s) \log{(1 - s)} + s(1 - s/2) \right] (1 - s) (3 + 5s) / 4s^2 ~ .
 \eea
The coefficient of the pion chiral log in Eq.~(\ref{eq:ff_SU3exp})  is in agreement with the one predicted by SU(2) ChPT \cite{Flynn:2008tg} both at $q^2 = 0$ and $q^2 = q_{max}^2$.
At $q^2 = 0$ the leading chiral log has the coefficient ($-3/4$), while close to $q^2 = q_{max}^2$, i.e.~for $s \simeq (1 - \sqrt{x})^2$, the function $T_0^{(1)}(s)$ contributes to the leading chiral log, obtaining for the scalar form factor $f_0(q_{max}^2)$ a final coefficient equal to ($-11/4$) in agreement with Ref.~\cite{Flynn:2008tg}.

As for the $q^2$-dependence, the results of the previous Section suggest that a simple pole Ansatz is able to reproduce quite well the lattice data for each gauge ensemble (see Eq.~(\ref{eq:zexpansion_ff})).
Note that the pole behavior is not predicted by either SU(3) or SU(2) ChPT at NLO \cite{Gasser:1984ux,Gasser:1984gg}, being a higher order effect.
Moreover, the slopes of the form factors $f_{+,0}(q^2)$ at $q^2 = 0$ (i.e.~the parameters $1 / M_{V,S}^2$ in Eq.~(\ref{eq:zexpansion_ff})) determined for each gauge ensemble exhibit an almost linear dependence on the light-quark mass $m_\ell$ and on the squared lattice spacing $a^2$.
In particular, as expected from the vector meson dominance model, the slope of the vector form factor extrapolated at the physical point is found to be consistent with the location of the physical $K^*(892)$ resonance.

Thus our Ansatz for the vector and scalar form factors, inspired by SU(2) ChPT, has the following form
\be
   \label{eq:ff_SU2}
   f_{+,0}(q^2) = \frac{ f_+^{SU(2)}(0) - F_+ \left[ x T_{+,0}^{(1)}(s) + T_{+,0}^{(2)}(s) \right] M_K^2 / (4 \pi f)^2}
                          {1 - q^2 [1 + P_{+,0} M_\pi^2 + D_{+,0} a^2 + K_{+,0}^{FSE}(L)] / \overline{M}_{V,S}^2} \left( 1 + A_{+,0} s \right) ~ , 
 \ee
where $f_+^{SU(2)}(0)$ is given by Eq.~(\ref{eq:SU2fit}), $\overline{M}_V$ is a parameter taken to be equal to the mass of the vector $K^*(892)$ resonance and $\overline{M}_S$ is left as a free parameter in the fit.
In Eq.~(\ref{eq:ff_SU2}) the quantity $K_{+,0}^{FSE}(L)$ is a phenomenological term that parameterizes the FSEs, which will be described soon, while $P_{+,0}$, $D_{+,0}$, $A_{+,0}$ and $F_+$ are free parameters.
In order to improve the fit quality we have inserted in the numerator of Eq.~(\ref{eq:ff_SU2}) an extra $q^2$-dependence through the factor ($1 + A_{+,0} s $).

\subsection{Finite size effects}
\label{sec:FSE}

\begin{figure}[htb!]
\begin{center}
\scalebox{0.85}{\includegraphics{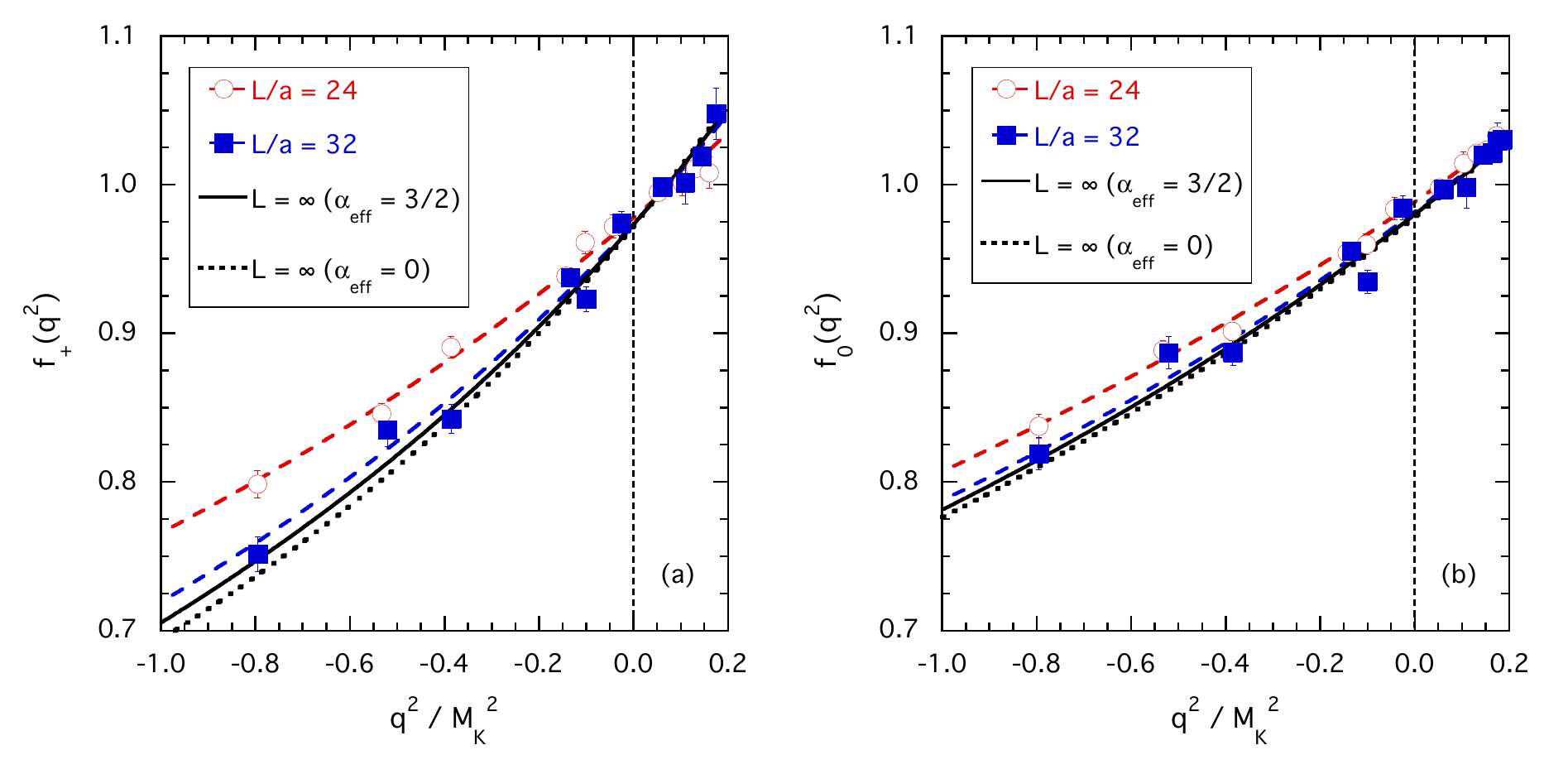}}
\end{center}
\vspace*{-1cm}
\caption{\it Results for the vector (a) and scalar (b) form factors as functions of $q^2 / M_K^2$ for the ensembles A40.24 and A40.32, which correspond to $\beta = 1.90$ and $M_\pi \simeq 300 \mev$ at two different lattice volumes $L / a = 24$ (red dots) and $L / a = 32$ (blue squares), respectively. The dashed lines represent the results of a simple pole fit to the data for each lattice volume separately. The solid and dotted lines correspond to the infinite volume predictions obtained using Eq.~(\ref{eq:KFSE}) for the slopes in the two cases $\alpha_{eff} = 0$ and $\alpha_{eff} = 3/2$.}
\label{fig:FSE}
\end{figure}

FSEs have been investigated by comparing the results obtained for the two ensembles A40.24 and A40.32, which share the same lattice spacing and pion masses at different lattice volumes. 
The comparison is illustrated in Fig.~\ref{fig:FSE}, which shows the presence of sizeable FSEs in the slopes of the form factors with a larger impact in the case of the vector form factor.
Instead FSEs do not exceed $\sim 1 $ standard deviation for the form factor at $q^2 = 0$.
We remind that FSEs are expected to be maximal for the ensemble A40.24.

The lattice data for the two volumes can be fitted by a simple pole Ansatz at each lattice volume separately (see dashed lines in Fig.~\ref{fig:FSE}).
This however cannot tell us anything about the infinite volume limit.
To this end we include FSEs on the the slopes of the form factors $f_{+,0}(q^2)$ at $q^2 = 0$ by introducing the following phenomenological parameterization
 \be
     \label{eq:KFSE}
     K_{+,0}^{FSE}(L) =  D_{+,0}^{FSE} M_\pi^2 \frac{e^{-M_\pi L}}{(M_\pi L)^{\alpha_{eff}}} ~ ,
 \ee
where $D_{+,0}^{FSE}$ are free parameters and $\alpha_{eff}$ is an effective power that controls the decrease of FSEs at large values of $M_\pi L$.
In the case $\alpha_{eff} = 3/2$ the term $e^{-M_\pi L} / (M_\pi L)^{3/2}$ is known to represent the leading FSE correction to the pion mass and decay constant for $M_\pi L \gg 1$  \cite{Gasser:1986vb}.

In Fig.~\ref{fig:FSE} we have given the predictions at infinite volume obtained using Eq.~(\ref{eq:KFSE}) for the slopes in the two cases $\alpha_{eff} = 0$ and $\alpha_{eff} = 3/2$.
It can be seen that in both cases the predictions at infinite volume are consistent with the data at the largest volume.
The spread between the results obtained using $\alpha_{eff} = 0$ and $\alpha_{eff} = 3/2$ will be used to estimate the systematic uncertainty related to FSEs.

\subsection{Modified $z$-expansion Ansatz}
\label{sec:zexp}

The second Ansatz for describing the $q^2$-, $m_\ell$- and $a$-dependencies of the form factors is a modified version of the $z$-expansion given by Eq.~(\ref{eq:zexpansion_ff}), in which we take into account, as in Eq.~(\ref{eq:ff_SU2}), that the slopes of the form factors at $q^2 = 0$ have an almost linear dependence on the (renormalized) light-quark mass $m_\ell$ and on the squared lattice spacing $a^2$.
We write
 \be
     \label{eq:zexpansion_mod}
     f_{+,0}(q^2 ) = \frac{f_+^{SU(3)}(0) + \tilde{A}_{+,0} ~ (z - z_0) \left[ 1 + (z + z_0) / 2 \right]} {1 - q^2 (1 + \tilde{P}_{+,0} M_\pi^2 + \tilde{D}_{+,0} a^2 + 
                             \tilde{K}_{+,0}^{FSE}) / \tilde{M}_{V,S}^2} ~ ,
 \ee
where $f_+^{SU(3)}(0)$ is given by Eqs.~(\ref{eq:SU3fit}-\ref{eq:Deltaf}), $\tilde{M}_V$ is a parameter taken to be equal to the mass of the vector $K^*(892)$ resonance and $\tilde{M}_S$ is left as a free parameter in the fit.

We include in our analysis the constraint coming from the CT theorem \cite{Callan:1966hu}, which states that the scalar form factor $f_0(q^2)$ at the (unphysical) CT point $q^2 = q_{CT}^2 = M_K^2 - M_\pi^2$, differs from the ratio of the leptonic decay constants $f_K / f_\pi$ by terms which are proportional to the light-quark mass, namely: $f_0(q^2 = M_K^2 - M_\pi^2) = f_K / f_\pi + {\cal{O}}(m_\ell)$.
Therefore, in the SU(2) chiral limit the scalar form factor $f_0(q^2)$ at $q^2 = q_{CT}^2 = q_{max}^2 = \overline{M}_K^2$ coincides with the ratio of the leptonic decay constants $\overline{f}_K / f$, where $\overline{f}_K$ and $\overline{M}_K$ are the SU(2) chiral limits of $f_K$ and $M_K$, respectively.
The CT theorem is therefore equivalent to impose on the parameters of Eqs.~(\ref{eq:ff_SU2}) and (\ref{eq:zexpansion_mod}) the following constraints
 \bea
     \label{eq:CT_constraints}
    F_+ \frac{1 + A_0}{1 - \overline{M}_K^2 / \overline{M}_S^2} & = & \frac{\overline{f}_K}{f} \qquad \qquad \mbox{[SU(2) ChPT inspired fit]} ~ , \nonumber \\[2mm]
    \frac{1 + \overline{f}_2 + \Delta_0 -2 \tilde{A}_0}{1 - \overline{M}_K^2 / \tilde{M}_S^2} & = & \frac{\overline{f}_K}{f} \qquad \qquad \mbox{[modified z-expansion fit]} ~ ,
 \eea
where $\overline{f}_2$ is the SU(2) chiral limit of Eq.~(\ref{eq:f2}) and the value for the ratio $\overline{f}_K / f$ is taken from the determination performed in Ref.~\cite{Carrasco:2014poa}.

Taking into account the constraints coming from the CT theorem the number of free parameters for the SU(2) and modified $z$-expansion fits is $11$ and $14$, respectively. 
Generally speaking, all these parameters are functions of the strange quark mass, which however is fixed in our analysis at its physical value.

\begin{figure}[htb!]
\begin{center}
\scalebox{0.65}{\includegraphics{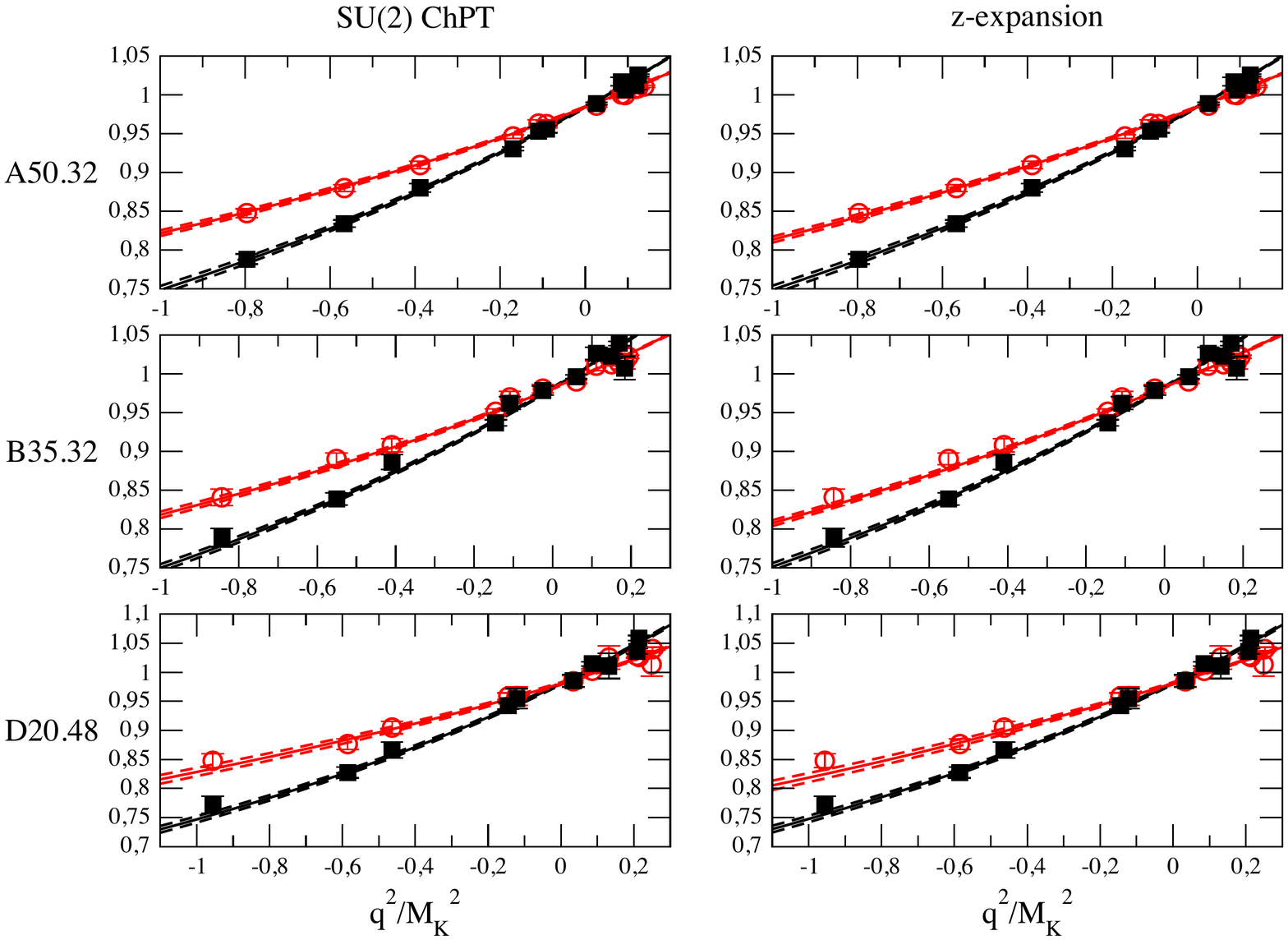}}
\end{center}
\vspace*{-0.8cm}
\caption{\it Vector (black squares) and scalar (red circles) form factors versus $q^2 / M_K^2$ for the ensembles A50.32, B35.32 and D20.48. The solid lines represent the results of the SU(2) ChPT (right panel) and the modified $z$-expansion (left panel) fits (see text). The dashed lines identify the uncertainties due to the statistical errors of the data and to the fitting procedure.}
\label{fig:combined_fit}
\end{figure}

The quality of both the SU(2) ChPT and the modified $z$-expansion fits to our data is quite good and it is illustrated in Fig.~\ref{fig:combined_fit} in the cases of the three ensembles A50.32, B35.32 and D20.48.

\begin{figure}[htb!]
\begin{center}
\scalebox{0.65}{\includegraphics{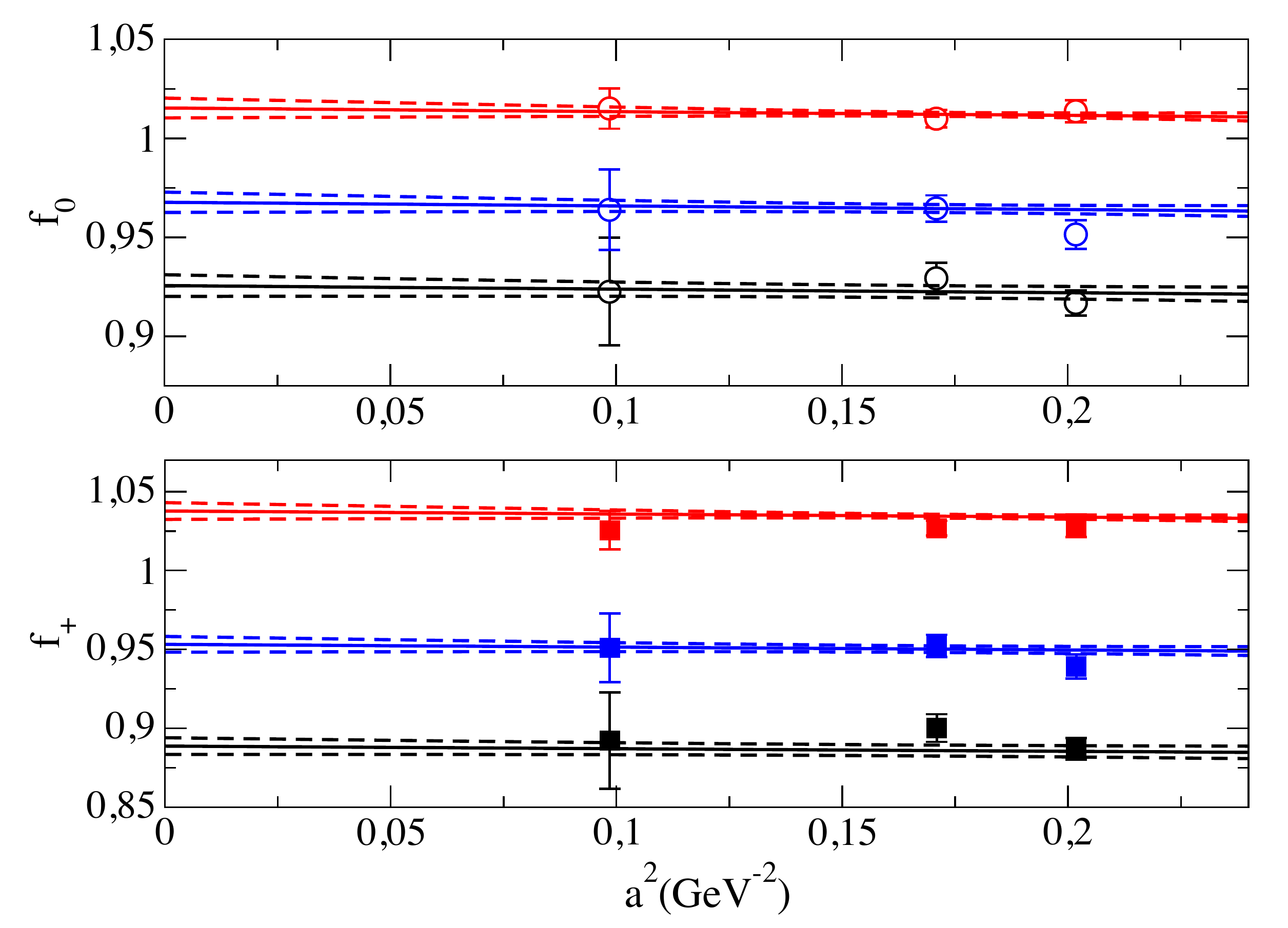}}
\end{center}
\vspace*{-0.8cm}
\caption{\it Scaling behavior of the vector (squares) and scalar (circles) form factors, interpolated at the references values $m_\ell = 20 \mev$, $m_s = 99.6 \mev$ and $q^2 / M_K^2 = -0.33, -0.11, 0.14$ corresponding to black, blue and red markers (or bottom to top), respectively. Solid and dashed lines represent the results of the SU(2) ChPT fit with its statistical uncertainty.}
\label{fig:scaling}
\end{figure}

The good scaling behavior of our lattice data for the vector and scalar form factors is illustrated in Fig.~\ref{fig:scaling}, where we have shown the results obtained after a small interpolations at the reference value $m_\ell = 20 \mev$ for the light-quark mass and at few reference values of the variable $q^2 / M_K^2$.
The data depends mildly on $a^2$ and the discretization effects do not exceed the permille level.

In Table \ref{tab:synthetic} we provide a set of synthetic data points with the corresponding total uncertainties representing our results for the vector and scalar form factors, extrapolated to the physical pion mass and to the continuum and infinite volume limits, for 12 selected values of $q^2$ in the range between $q^2 = - q_{max}^2$ and $q^2 = q_{max}^2$ with $q_{max}^2 \simeq 0.129 \gev^2$.

\begin{table}[htb!]
\begin{center}
\begin{tabular}{||c||c|c||}
\hline
$q^2 ~ (\mbox{GeV}^2)$ & $f_+(q^2)$ & $f_0(q^2)$  \\ \hline \hline
 $-0.1290$ & ~ $0.8288 ~ (81)$ ~ & ~ $0.8911 ~ (100)$ ~   \\ \hline 
 $-0.0968$ & ~ $0.8603 ~ (72)$ ~ & ~ $0.9097 ~ (86)$ ~   \\ \hline 
 $-0.0645$ & ~ $0.8943 ~ (62)$ ~ & ~ $0.9290 ~ (72)$ ~   \\ \hline 
 $-0.0323$ & ~ $0.9311 ~ (53)$ ~ & ~ $0.9494 ~ (59)$ ~   \\ \hline 
 $0.0$ & ~ $0.9711 ~ (47)$ ~ & ~ $0.9711 ~ (47)$ ~ \\ \hline 
 $+0.0184$ & ~ $0.9954 ~ (47)$ ~ & ~ $0.9840 ~ (42)$ ~  \\ \hline 
 $+0.0369$ & ~ $1.0211 ~ (49)$ ~ & ~ $0.9976 ~ (37)$ ~  \\ \hline 
 $+0.0553$ & ~ $1.0480 ~ (55)$ ~ & ~ $1.0117 ~ (35)$ ~  \\ \hline 
 $+0.0737$ & ~ $1.0764 ~ (64)$ ~ & ~ $1.0264 ~ (36)$ ~  \\ \hline 
 $+0.0922$ & ~ $1.1062 ~ (76)$ ~ & ~ $1.0419 ~ (42)$ ~  \\ \hline 
 $+0.1106$ & ~ $1.1378 ~ (90)$ ~ & ~ $1.0581 ~ (52)$ ~  \\ \hline 
 $+0.1290$ & ~ $1.1711 ~ (106)$ ~ & ~ $1.0753 ~ (67)$ ~  \\ \hline 
\end{tabular}
\end{center}
\caption{\it Synthetic data points representing our results for the vector and scalar form factors extrapolated to the physical pion point and in the continuum and infinite volume limits for twelve selected values of $q^2$ in the range between $q^2 = - q_{max}^2$ and $q^2 = q_{max}^2$, where $q_{max}^2 \simeq 0.129 \gev^2$ is the physical kinematical end-point. The errors correspond to the combination in quadrature of both statistical and systematic errors (see text). }
\label{tab:synthetic}
\end{table}

The uncertainties correspond to the combination in quadrature of both statistical and systematic errors.
These errors take into account (always in quadrature) the uncertainties induced by: ~ i) the statistical noise and the fitting procedure; ii) the errors present in the determination of the input parameters used in our analysis, namely the values of the average $u/d$ quark mass $m_{ud}$, the lattice spacing $a$ and the SU(2) ChPT LEC's $f$ and $B$, determined in Ref.~\cite{Carrasco:2014cwa}; ~ iii) the chiral extrapolation, evaluated as half of the difference of the results obtained using the SU(2) ChPT and the modified z-expansion fits, ~ iv) FSEs, evaluated by comparing the results obtained using the FSE factor (\ref{eq:KFSE}) with $\alpha_{eff} =0$ and $\alpha_{eff} = 3/2$, and ~ v) discretization effects, calculated by comparing with the results obtained including in the denominator of Eqs.~(\ref{eq:ff_SU2}) and (\ref{eq:zexpansion_mod}) a term proportional to $(a/r_0)^4$ and adopting for its coefficient a prior distribution equal to $0 \pm 3$.

\subsection{Dispersive parameterization}
\label{sec:dispersive}

In order to fit the $q^2$-dependence of our synthetic data points we have used the same parameterization adopted in the analyses of the experimental data \cite{FlaviaNet,Moulson:2014cra}, namely the dispersive parameterization of Ref.~\cite{Bernard:2009zm} which reads
 \bea
       \label{eq:f+_dispersive}
       f_+^{disp.}(q^2) & = & f_+(0) ~ e^{\frac{q^2}{M_\pi^2} \left[ \Lambda_+ + H(q^2) \right]} ~ , \\[2mm]
       \label{eq:f0_dispersive}
       f_0^{disp.}(q^2) & = & f_+(0) ~ e^{\frac{q^2}{q_{CT}^2} \left[ \rm{log}(C) - G(q^2) \right]} ~ , 
  \eea
where the dispersive functions $H(q^2)$ and $G(q^2)$ are given by
 \bea
        \label{eq:Hq2}
       H(q^2) & \equiv & \frac{M_\pi^2 q^2}{\pi} \int_{q_{cut}^2}^\infty ds ~ \frac{\phi_+(s)}{s^2 (s - q^2 -i \varepsilon)} ~ , \\
       \label{eq:Gq2}
       G(q^2) & \equiv & \frac{q_{CT}^2 (q_{CT}^2 - q^2)}{\pi} \int_{q_{cut}^2}^\infty ds ~ \frac{\phi_0(s)}{s (s - q_{CT}^2) (s - q^2 - i \varepsilon)}  
  \eea
with $q_{cut}^2 \equiv (M_K + M_\pi)^2$.
The parameters $\Lambda_+$ and $C$ represent the slope of the vector form factor $f_+(q^2)$ at $q^2 = 0$ (in units of $M_\pi^2$) and the value of the scalar form factor $f_0(q^2)$ at the CT point $q^2 = q_{CT}^2 \equiv M_K^2 - M_\pi^2$ divided by $f_+(0)$, respectively.
In Eqs.~(\ref{eq:Hq2}-\ref{eq:Gq2}) the quantities $\phi_+$ and $\phi_0$ can be identified in the elastic region with the $P$-wave and $S$-wave phase shifts of the $(K \pi)_{I=1/2}$ scattering.
The functions $H(q^2)$ and $G(q^2)$ have been estimated numerically in Ref.~\cite{Bernard:2009zm} with a $\simeq 10 \%$ accuracy and their contributions do not exceed $\simeq 20 \%$ of the value of $\Lambda_+$ and $\rm{log}(C)$, respectively.

After performing a Taylor expansion of the form factors (\ref{eq:f+_dispersive}-\ref{eq:f0_dispersive}), viz.
 \be
      \label{eq:Taylor}
      f_{+,0}^{disp.}(q^2) = f_+(0) ~ \left\{ 1 + \lambda_{+,0}^{'} \frac{q^2}{M_\pi^2} + \frac{1}{2} \lambda_{+,0}^{''} \left( \frac{q^2}{M_\pi^2} \right)^2 + 
                                        \frac{1}{6} \lambda_{+,0}^{'''} \left( \frac{q^2}{M_\pi^2} \right)^3 + \dots \right\} ~ ,
  \ee
one has \cite{Bernard:2009zm}
 \bea
       \label{eq:lambda+_parms}
       \lambda_+^{'} & = & \Lambda_+ ~ , \\
       \lambda_+^{''} & = & ( \lambda_+^{'} )^2 + 5.79 (97) \cdot 10^{-4} ~ , \nonumber \\
       \lambda_+^{'''} & = & ( \lambda_+^{'} )^3 + 5.79 (97) \cdot 10^{-4} ~ 3 \lambda_+^{'} + 2.99 (21) \cdot 10^{-5} \nonumber
 \eea
and
 \bea
       \label{eq:lambda0_parms}
       \lambda_0^{'} & = & \frac{M_\pi^2}{q_{CT}^2} \left[ \rm{log}(C) - 0.0398 (44) \right] ~ , \\[2mm]
       \lambda_0^{''} & = & ( \lambda_0^{'} )^2 + 4.16 (56) \cdot 10^{-4} ~ , \nonumber \\[2mm]
       \lambda_0^{'''} & = & ( \lambda_0^{'} )^3 + 4.16 (56) \cdot 10^{-4} ~ 3 \lambda_0^{'} + 2.72 (21) \cdot 10^{-5}  ~ . \nonumber
 \eea

We have applied the dispersive parameterization (\ref{eq:Taylor}-\ref{eq:lambda0_parms}) to the synthetic data points of Table \ref{tab:synthetic}, having three parameters to be determined, namely: the form factor $f_+(0)$, the vector slope $\Lambda_+$ and the logarithm of the scalar form factor at the CT point divided by $f_+(0)$, $\rm{log}(C)$.
Using our bootstrap samples, which allow to take properly into account the correlations among the synthetic data points, we have obtained
 \bea
       \label{eq:f+(0)_final}
       f_+(0) & = & 0.9709 ~ (44)_{stat+fit} ~ (11)_{input} ~ (9)_{syst} = 0.9709 ~ (46) ~ , \\
       \label{eq:lambda+_final}
       \Lambda_+ & = & 24.22 ~ (1.05)_{stat+fit} ~ (0.44)_{input} ~ (0.25)_{syst} \cdot 10^{-3} = 24.22 ~ (1.16) \cdot 10^{-3} ~ , \\
       \label{eq:logC_final}
       \rm{log}(C) & = & 0.1998 ~ (133)_{stat+fit} ~ (36)_{input} ~ (15)_{syst} = 0.1998 ~ (138) ~ ,
 \eea
where all the uncertainties are combined in quadrature in the final error and the systematic error takes into account the uncertainties related to the chiral extrapolation, finite volume and discretization effects.
The error budgets are given in Table \ref{tab:budget} and the correlation coefficients are equal to
 \bea
        \label{eq:rho12}
        \rho[f_+(0), \Lambda_+] & = & -0.228 ~ , \\
        \label{eq:rho13}
        \rho[f_+(0), \rm{log}(C) ] & = & -0.719 ~ , \\
        \label{eq:rho23}
        \rho[\Lambda_+, \rm{log}(C) ] & = & 0.376 ~ .
 \eea

\begin{table}[htb!]
\begin{center}
\begin{tabular}{||c||c|c|c||}
\hline
 uncertainty & $f_+(0)$ & $\Lambda_+ \cdot 10^3$ & ${\rm log}(C)$  \\ \hline \hline
 statistics + fitting procedure                         & $0.0044$ & ~ $1.05$ ~ & ~ $0.0133$ ~ \\ \hline
 input parameters \cite{Carrasco:2014cwa}  & $0.0011$ & ~ $0.44$ ~ & ~ $0.0036$ ~ \\ \hline 
 chiral extrapolation                                       & $0.0009$ & ~ $0.17$ ~ & ~ $0.0015$ ~ \\ \hline 
 $a^2 \to 0$ extrapolation                              & $0.0003$ & ~ $0.13$ ~ & ~ $0.0002$ ~ \\ \hline 
 finite volume correction                                & $0.0001$ & ~ $0.12$ ~ & ~ $0.0002$ ~ \\ \hline
 total                                                              & $0.0046$ & ~ $1.16$ ~ & ~ $0.0138$ ~ \\ \hline \hline
\end{tabular}
\end{center}
\caption{\it Sources of uncertainty in the final results (\ref{eq:f+(0)_final}-\ref{eq:logC_final}) for the vector form factor $f_+(0)$ and the parameters $\Lambda_+$ and ${\rm log}(C)$ of the dispersive fit (\ref{eq:Taylor}-\ref{eq:lambda0_parms}) performed using the synthetic data of Table \ref{tab:synthetic} at the physical point.}
\label{tab:budget}
\end{table}

The result (\ref{eq:f+(0)_final}) for $f_+(0)$ is in nice agreement with the one of Eq.~(\ref{eq:f+0_final1}) obtained using the first strategy of our analysis (see Section \ref{sec:first_strategy}).
Being slightly more precise, we take it as our final result for $f_+(0)$.
Eq.~(\ref{eq:f+(0)_final}) is compatible with the current FLAG average $f_+(0) = 0.9661 (32)$ at $N_f = 2+1$ \cite{FLAG2}, based on the results from FNAL/MILC \cite{Bazavov:2012cd} and RBC/UKQCD \cite{Boyle:2013gsa}. 
It is also consistent with the recent determinations $f_+(0) = 0.9704 (24)_{stat} (22)_{syst} = 0.9704 (32)$ from FNAL/MILC at $N_f = 2+1+1$ \cite{Bazavov:2013maa} and $f_+(0) = 0.9685 (34)_{stat} (14)_{syst} = 0.9685 (37)$ from RBC/UKQCD at $N_f = 2+1$ \cite{Boyle:2015hfa}.

\begin{figure}[htb!]
\begin{center}
\scalebox{0.85}{\includegraphics{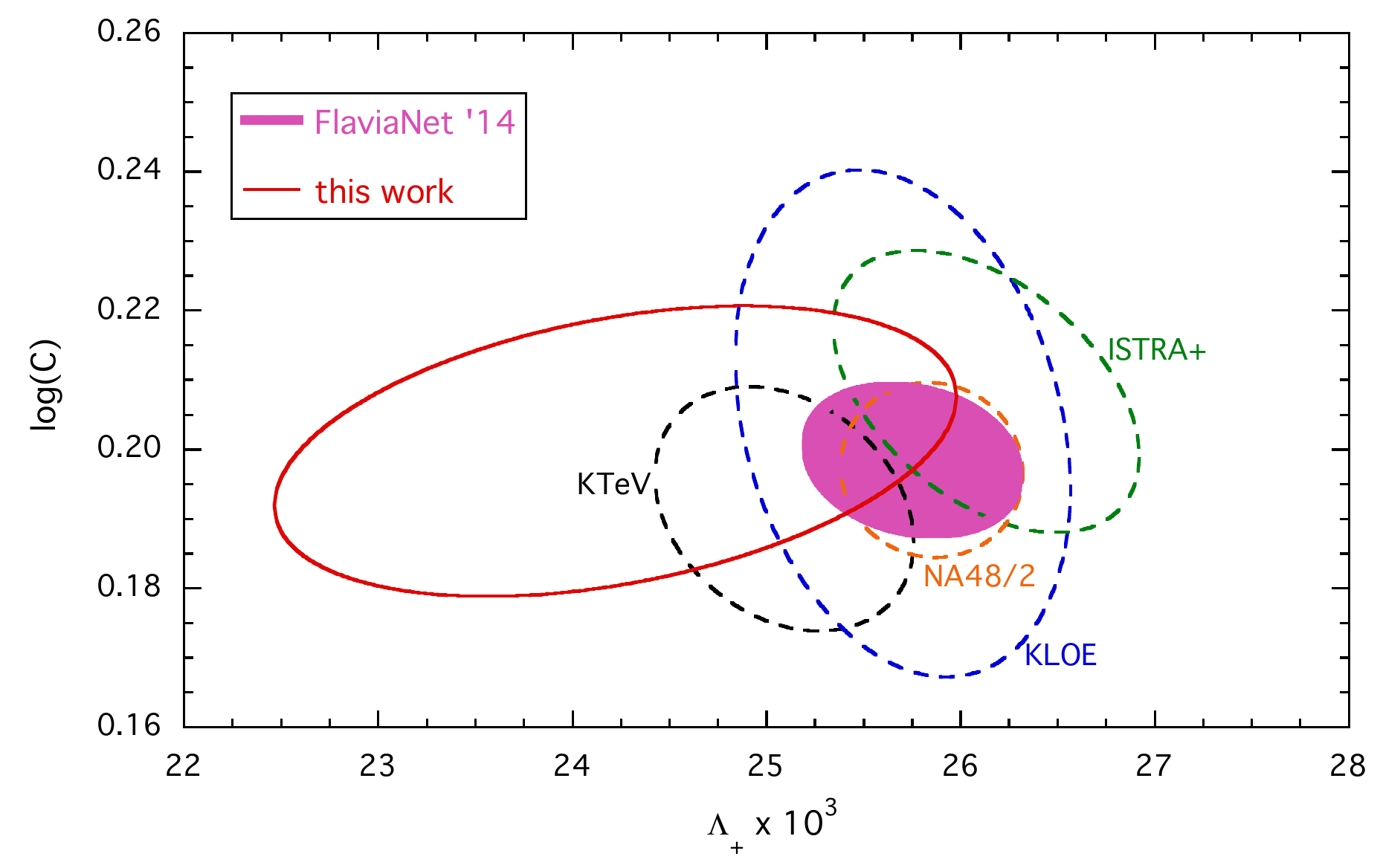}}
\end{center}
\vspace*{-1cm}
\caption{\it Comparison of the information for the dispersive parameters $\Lambda_+$ and $\rm{log}(C)$ obtained in this work (solid ellipse) with the results of the $K_{\ell 3}$ experiments KTeV, KLOE, NA48/2 and ISTRA+ (dashed ellipses), taken from Refs.~\cite{FlaviaNet,Moulson:2014cra}, and with the updated FlaviaNet average of Ref.~\cite{Moulson:2014cra} (full ellipse). All the ellipses represent contours corresponding to a $68\%$ likelihood.}
\label{fig:contours}
\end{figure}

Finally, our results (\ref{eq:lambda+_final}-\ref{eq:logC_final}) compare positively with the corresponding updated FlaviaNet average of experimental results obtained in Ref.~\cite{Moulson:2014cra}:
 \bea
     \label{eq:lambda+_exp}
     \Lambda_+^{exp} & = & 25.75 ~ (36) \cdot 10^{-3} ~ , \\
     \label{eq:logC_exp}
     \rm{log}(C)^{exp} & = & 0.1985 ~ (70) ~ .
 \eea
The agreement is illustrated in Fig.~\ref{fig:contours}, where the ($68\%$ likelihood) contour corresponding to our results (\ref{eq:lambda+_final}-\ref{eq:logC_final}), taking into account the correlation coefficient (\ref{eq:rho23}), is compared with the corresponding information coming from various $K_{\ell 3}$ experiments (taken from Refs.~\cite{FlaviaNet,Moulson:2014cra}) and with the updated FlaviaNet average of Ref.~\cite{Moulson:2014cra}.

\begin{figure}[htb!]
\begin{center}
\scalebox{0.65}{\includegraphics{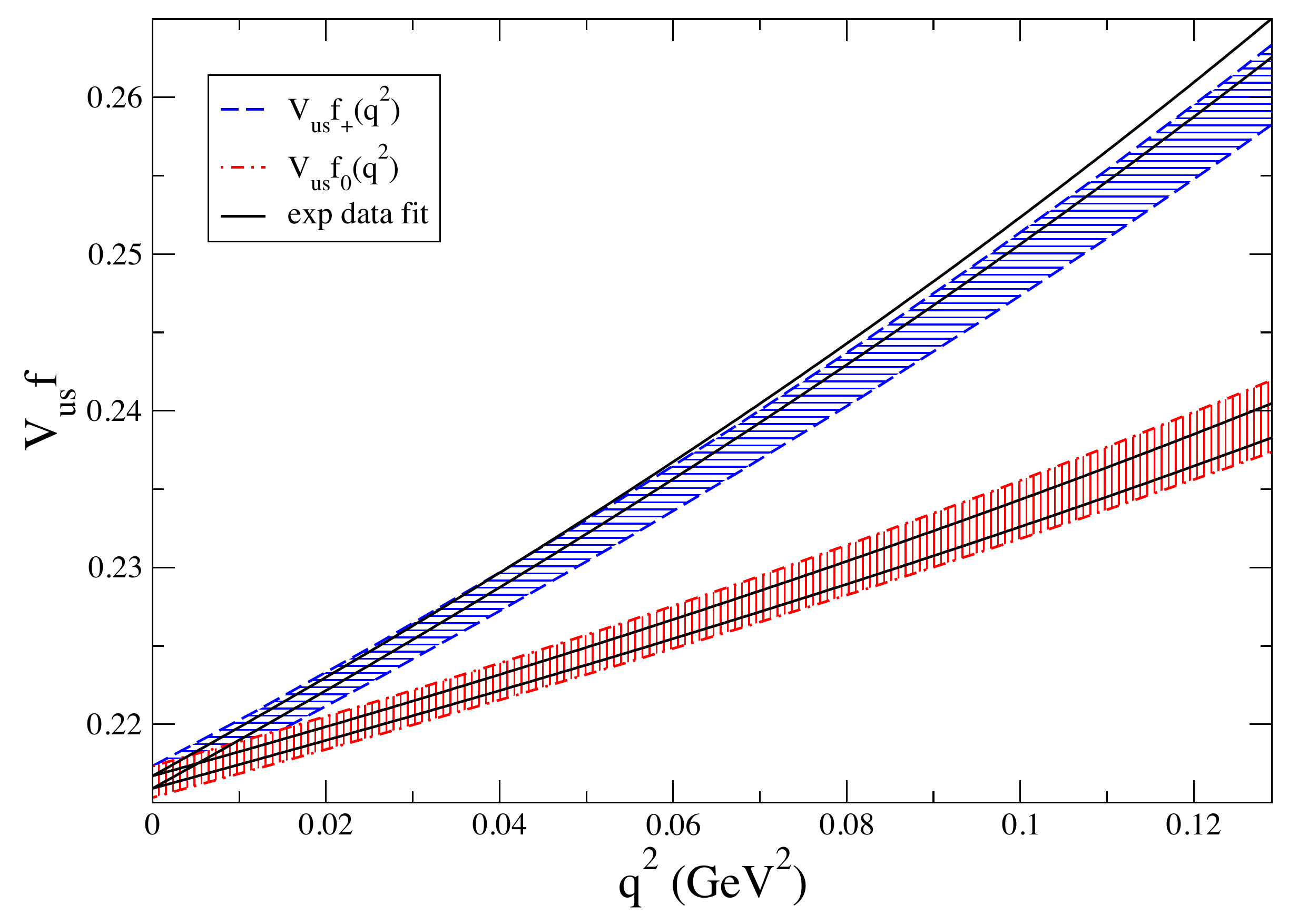}}
\end{center}
\vspace*{-1cm}
\caption{\it Results for the vector (blue area) and scalar (red area) form factors, obtained at the physical point including both statistical and systematic uncertainties (see text), multiplied by $|V_{us}| = 0.2230$ (see Eq.~(\ref{eq:Vus_Kl3})) versus $q^2$ in the range between $q^2 = 0$ and the physical kinematical end-point $q^2 = q_{max}^2 \simeq 0.129 \gev^2$. The black solid lines represent the results of the dispersive fit of the experimental data performed in Ref.~\cite{Moulson:2014cra}.}
\label{fig:f0fp_physical_experiment}
\end{figure}

The momentum dependence of the vector and scalar form factors obtained from our dispersive fits can be inferred from Fig.~\ref{fig:f0fp_physical_experiment}, where the results for $f_{+,0}(q^2)$ are multiplied by $|V_{us}| = 0.2230$ (see later Eq.~(\ref{eq:Vus_Kl3})) and compared with those obtained in Ref.~\cite{Moulson:2014cra} by applying the dispersive parameterization of Ref.~\cite{Bernard:2006gy,Bernard:2009zm} to the experimental data.

The agreement with the experimental data shown in Figs.~\ref{fig:contours}-\ref{fig:f0fp_physical_experiment} is remarkable in spite of the larger uncertainties affecting the theoretical results.
This provides a strong motivation for future investigations of the semileptonic vector and scalar form factors, which will improve the precision of the theoretical predictions not only at the particular kinematical point $q^2 =0$, but in the full $q^2$-range covered by the experiments, obtaining in this way a more stringent test of the SM in $K_{\ell 3}$ decays. 

In order to allow a direct use of the synthetic data points without using our bootstrap samples, we have calculated the covariance matrix among the synthetic data points of Table \ref{tab:synthetic}.
Its dimension is $23 \times 23$ and it cannot be easily given in tables.
We can therefore supply it upon request.

\section{Evaluation of $|V_{us}|$}
\label{sec:Vus}

Combining our final result (\ref{eq:f+(0)_final}) for the vector form factor $f_+(0)$ with the experimental value of $|V_{us}| f_+(0) = 0.2165 (4)$ from Ref.~\cite{Moulson:2014cra} we can estimate the CKM matrix element $|V_{us}|$, obtaining
 \be
    \label{eq:Vus_Kl3}
    |V_{us}| = 0.2230 ~ (4)_{exp} ~ (11)_{f_+(0)}  = 0.2230 ~ (11) ~ \qquad \mbox{from $K_{\ell 3}$ decays} ~ , 
 \ee 
where the errors come from experiments and from the lattice computation, respectively.

The result (\ref{eq:Vus_Kl3}) may be compared with the determination of $|V_{us}|$ coming from the ratio of the kaon and pion leptonic decay constants $f_{K^+} /f_{\pi^+} = 1.184 (16)$, obtained in Ref.~\cite{Carrasco:2014poa} using the same ETMC gauge configurations for the lattice calculations. 
Taking the updated value of $|V_{ud}| = 0.97417 (21)$ from the superallowed nuclear $\beta-$decay \cite{Hardy:2014qxa} and the updated experimental value of $|V_{us} / V_{ud}| f_{K^+} / f_{\pi^+} = 0.2760 (4)$ from Ref.~\cite{Moulson:2014cra} one gets
 \be
     \label{eq:Vus_Kl2}
     |V_{us}| = 0.2271 ~ (29) ~ \qquad \qquad \qquad \mbox{from $K_{\ell 2}$ decays} ~ ,
 \ee
which is in slight tension ($\sim 1.3$ standard deviations) with the result (\ref{eq:Vus_Kl3}) obtained from $K_{\ell 3}$ decays\footnote{The slight tension among the results (\ref{eq:Vus_Kl3}-\ref{eq:Vus_Kl2}) is not significantly changed by the small correlation coefficient occurring between our result (\ref{eq:f+(0)_final}) for $f_+(0)$ and the ETMC value of $f_{K^+} /f_{\pi^+}$ obtained in Ref.~\cite{Carrasco:2014poa}.}. 
We can also use our results (\ref{eq:Vus_Kl3}-\ref{eq:Vus_Kl2}) to test the unitarity of the first row of the CKM matrix using the updated PDG value of $|V_{ub}| = 0.00413 (49)$ from $B$-meson decays \cite{PDG}.
One obtains
 \bea
        \label{eq:utest}
        |V_{ud}|^2 + |V_{us}|^2 + |V_{ub}|^2 & = & 0.99875 ~ (64) ~ \qquad \mbox{from $K_{\ell 3}$ decays} ~ , \nonumber \\[2mm]
        |V_{ud}|^2 + |V_{us}|^2 + |V_{ub}|^2 & = & 1.0008 ~ (14) ~ \qquad \mbox{~ from $K_{\ell 2}$ decays} ~ ,
 \eea
which test the first-row unitarity at the permille level for the $K_{\ell 2}$ decays and even below for the $K_{\ell 3}$ modes.
In the latter case a slight tension with unitarity at the level of $\sim 2$ standard deviations is observed.

\section{Conclusions}
\label{sec:conclusions}

We present a new lattice QCD determination of the vector and scalar form factors of the semileptonic $K \to \pi \ell \nu$ decay which are relevant for the
extraction of the CKM matrix element $|V_{us}|$ from experimental data.

Our results are based on the gauge configurations produced by the European Twisted Mass Collaboration with $N_f = 2+1+1$ dynamical fermions, which include in the sea, besides two light mass degenerate quarks, also the strange and the charm quarks.
Using data simulated at three different values of the lattice spacing and with pion masses as small as $210$ MeV, our final result for the vector form factor at zero momentum transfer is 
 \be
     f_+(0) = 0.9709 ~ (44)_{stat+fit} ~ (11)_{input} ~ (9)_{syst} = 0.9709 ~ (46) ~ ,
 \ee
where all the uncertainties are combined in quadrature in the final error. 

Using the latest experimental value of $f_+(0) |V_{us}|$ from $K_{\ell 3}$ decays \cite{Moulson:2014cra}, we obtain
 \be
      |V_{us}| = 0.2230 ~ (11) ~ .
 \ee
This allows to test the unitarity constraint of the Standard Model below the permille level once the determination of $V_{ud}$ from superallowed nuclear $\beta$ decays is adopted, namely
 \be
        |V_{ud}|^2 + |V_{us}|^2 + |V_{ub}|^2  =  0.99875 ~ (64) ~ ,
 \ee
which highlights also a slight tension with unitarity at the level of $\sim 2$ standard deviations.

Besides $f_+(0)$ we have determined from our lattice data the values of the parameters $\Lambda_+$ and $\rm{log}(C)$ appearing in the dispersive parameterization (\ref{eq:Taylor}-\ref{eq:lambda0_parms}), which is adopted also to describe the experimental data \cite{FlaviaNet,Moulson:2014cra}.
Our results are given in Eqs.~(\ref{eq:lambda+_final}-\ref{eq:logC_final}) and compare positively with the corresponding latest experimental results (\ref{eq:lambda+_exp}-\ref{eq:logC_exp}) from Ref.~\cite{Moulson:2014cra}.
The consistency with the information coming from various $K_{\ell 3}$ experiments \cite{FlaviaNet,Moulson:2014cra} is remarkable, as shown in Fig.~\ref{fig:contours}.

We have also presented our results for the semileptonic scalar $f_0(q^2)$ and vector $f_+(q^2)$ form factors in the whole range of values of the squared four-momentum transfer $q^2$ measured in $K_{\ell 3}$ decays, obtaining a very good agreement with the momentum dependence of the experimental data, as illustrated in Fig.~\ref{fig:f0fp_physical_experiment}.

Our findings represent a strong motivation for future investigations of the semileptonic vector and scalar form factors, which will improve the precision of the theoretical predictions not only at the particular kinematical point $q^2 = 0$, but in the full $q^2$-range covered by the experiments, obtaining in this way a more stringent test of the Standard Model in $K_{\ell 3}$ decays.

Finally we have provided a set of synthetic data points representing our results for the vector and scalar form factors at the physical point for selected values of $q^2$ in the range between $q^2 = 0$ and the physical end-point $q^2 \simeq 0.129 \gev^2$ (including the covariance matrix for the data at different values of $q^2$, which is available upon request).

\section*{Acknowledgements}

\noindent We thank our colleagues of the ETM Collaboration for fruitful discussions.

\noindent We acknowledge the CPU time provided by the PRACE Research Infrastructure under the project PRA027 ``QCD Simulations for Flavor Physics in the Standard Model and Beyond'' on the JUGENE BG/P system at the J\"ulich SuperComputing Center (Germany), and by the agreement between INFN and CINECA under the specific initiatives INFN-RM123 and INFN-LQCD123 on the BG/Q system Fermi at CINECA (Italy).

\noindent V.~L., S.~S.~and C.~T.~thank MIUR (Italy) for partial support under Contract No. PRIN 2010-2011.

\noindent L.~R.~thanks INFN (Italy) for the support under the SUper MAssive (SUMA) computing project (https://web2.infn.it/SUMA).



\begin{thebibliography}{99}

\bibitem{CKM} 
N. Cabibbo, 
Phys.\ Rev.\ Lett.\  {\bf 10} (1963) 531.

M. Kobayashi and T. Maskawa,  
Prog.\ Theor.\ Phys.\ {\bf 49} (1973) 652.

\bibitem{FlaviaNet}
  M.~Antonelli {\it et al.} [FlaviaNet Working Group on Kaon Decays Collaboration],
  Eur.\ Phys.\ J.\ C {\bf 69} (2010) 399
  [arXiv:1005.2323 [hep-ph]].

\bibitem{FLAG1}
G.~Colangelo {\it et al.},
  Eur.\ Phys.\ J.\ C {\bf 71} (2011) 1695
  [arXiv:1011.4408 [hep-lat]].

\bibitem{FLAG2}
  S.~Aoki {\it et al.},
  Eur.\ Phys.\ J.\ C {\bf 74} (2014) 9,  2890
  [arXiv:1310.8555 [hep-lat]].

\bibitem{Becirevic:2004ya}
  D.~Becirevic {\it et al.},
  Nucl.\ Phys.\  B {\bf 705} (2005) 339
  [arXiv:hep-ph/0403217].

\bibitem{Bazavov:2013maa}
  A.~Bazavov {\it et al.},
  Phys.\ Rev.\ Lett.\  {\bf 112} (2014) 11,  112001
  [arXiv:1312.1228 [hep-ph]].

\bibitem{Boyle:2015hfa}
  P.~A.~Boyle {\it et al.} [RBC/UKQCD Collaboration],
  JHEP {\bf 1506} (2015) 164
  [arXiv:1504.01692 [hep-lat]].

\bibitem{Baron:2010bv}
  R.~Baron {\it et al.} [ETM Collaboration],
  JHEP {\bf 1006} (2010) 111
  [arXiv:1004.5284 [hep-lat]].

\bibitem{Baron:2011sf}
  R.~Baron {\it et al.}  [ETM Collaboration],
  PoS LATTICE {\bf 2010} (2010) 123
  [arXiv:1101.0518 [hep-lat]].

\bibitem{Moulson:2014cra}
  M.~Moulson,
  arXiv:1411.5252 [hep-ex].

\bibitem{Carrasco:2014pta}
  L.~Riggio {\it et al.} [ETM Collaboration],
  PoS LATTICE {\bf 2014} (2014) 387
  [arXiv:1411.1201 [hep-lat]].

\bibitem{Carrasco:2014uda}
  N.~Carrasco, P.~Lami, V.~Lubicz, E.~Picca, L.~Riggio, S.~Simula and C.~Tarantino,
  arXiv:1410.7159 [hep-lat].

\bibitem{Carrasco:2015wzu}
  N.~Carrasco, P.~Lami, V.~Lubicz, L.~Riggio and S.~Simula,
  arXiv:1511.04880 [hep-lat].

 \bibitem{Carrasco:2014cwa}
  N.~Carrasco {\it et al.},
  Nucl.\ Phys.\ B {\bf 887} (2014) 19
  [arXiv:1403.4504 [hep-lat]].

\bibitem{Carrasco:2014poa}
  N.~Carrasco {\it et al.},
  Phys.\ Rev.\ D {\bf 91} (2015) 5,  054507
  [arXiv:1411.7908 [hep-lat]].

\bibitem{Iwasaki:1985we}
  Y.~Iwasaki,
  Nucl.\ Phys.\ B {\bf 258} (1985) 141.

\bibitem{Frezzotti:2003xj}
  R.~Frezzotti and G.C.~Rossi,
  Nucl.\ Phys.\ Proc.\ Suppl.\  {\bf 128} (2004) 193
  [hep-lat/0311008].
  
\bibitem{Frezzotti:2003ni}
  R.~Frezzotti and G.C.~Rossi,
  JHEP {\bf 0408} (2004) 007
  [hep-lat/0306014].
 
 \bibitem{Frezzotti:2004wz}
  R.~Frezzotti and G.C.~Rossi,
  JHEP {\bf 0410} (2004) 070
  [hep-lat/0407002].
 
\bibitem{Osterwalder:1977pc}
  K.~Osterwalder and E.~Seiler,
  Annals Phys.\  {\bf 110} (1978) 440.

\bibitem{Sommer:1993ce}
  R.~Sommer,
  Nucl.\ Phys.\ B {\bf 411} (1994) 839
  [hep-lat/9310022].
 
\bibitem{Constantinou:2010gr}
  M.~Constantinou {\it et al.}  [ETM Collaboration],
  JHEP {\bf 1008} (2010) 068
  [arXiv:1004.1115 [hep-lat]].

\bibitem{Hardy:2014qxa}
  J.~C.~Hardy and I.~S.~Towner,
  Phys.\ Rev.\ C {\bf 91} (2015) 2,  025501
  [arXiv:1411.5987 [nucl-ex]].

\bibitem{PDG}
  K.~A.~Olive {\it et al.}  [Particle Data Group Collaboration],
  Chin.\ Phys.\ C {\bf 38} (2014) 090001.

\bibitem{Callan:1966hu}
  C.~G.~Callan and S.~B.~Treiman,
  Phys.\ Rev.\ Lett.\  {\bf 16} (1966) 153.

\bibitem{McNeile:2006bz}
  C.~McNeile and C.~Michael [UKQCD Collaboration],
  Phys.\ Rev.\  D {\bf 73} (2006) 074506
  [hep-lat/0603007].

\bibitem{Frezzotti:2008dr}
  R.~Frezzotti {\it et al.} [ETM Collaboration],
  Phys.\ Rev.\ D {\bf 79} (2009) 074506
  [arXiv:0812.4042 [hep-lat]].

\bibitem{Bedaque:2004kc}
  P.~F.~Bedaque,
  Phys.\ Lett.\ B {\bf 593} (2004) 82
  [nucl-th/0402051].
  
\bibitem{deDivitiis:2004kq}
  G.~M.~de Divitiis, R.~Petronzio and N.~Tantalo,
  Phys.\ Lett.\ B {\bf 595} (2004) 408
  [hep-lat/0405002].
 
\bibitem{Guadagnoli:2005be}
  D.~Guadagnoli, F.~Mescia and S.~Simula,
  Phys.\ Rev.\ D {\bf 73} (2006) 114504
  [hep-lat/0512020].

\bibitem{Flynn:2005in}
  J.~M.~Flynn {\it et al.} [UKQCD Collaboration],
  Phys.\ Lett.\ B {\bf 632} (2006) 313
  [hep-lat/0506016].

\bibitem{Boyle:2007wg}
  P.~A.~Boyle, J.~M.~Flynn, A.~Juttner, C.~T.~Sachrajda and J.~M.~Zanotti,
  JHEP {\bf 0705} (2007) 016
  [hep-lat/0703005 [HEP-LAT]].

\bibitem{Boyle:2010bh}
  P.~A.~Boyle {\it et al.} [RBC-UKQCD Collaboration],
  Eur.\ Phys.\ J.\ C {\bf 69} (2010) 159
  [arXiv:1004.0886 [hep-lat]].
 
\bibitem{Lubicz:2009ht}
  V.~Lubicz {\it et al.} [ETM Collaboration],
  Phys.\ Rev.\ D {\bf 80} (2009) 111502
  [arXiv:0906.4728 [hep-lat]].

\bibitem{Sachrajda:2004mi}
  C.~T.~Sachrajda and G.~Villadoro,
  Phys.\ Lett.\ B {\bf 609} (2005) 73
  [hep-lat/0411033].
 
 \bibitem{Bedaque:2004ax}
  P.~F.~Bedaque and J.~W.~Chen,
  Phys.\ Lett.\ B {\bf 616} (2005) 208
  [hep-lat/0412023].
     
\bibitem{Bourrely:2008za} 
  C.~Bourrely, I.~Caprini and L.~Lellouch,
  Phys.\ Rev.\ D {\bf 79}, 013008 (2009)
  [Erratum-ibid.\ D {\bf 82}, 099902 (2010)]
  [arXiv:0807.2722 [hep-ph]].
 
\bibitem{Flynn:2008tg}
  J.~M.~Flynn {\it et al.}  [RBC and UKQCD Collaborations],
  Nucl.\ Phys.\ B {\bf 812} (2009) 64
  [arXiv:0809.1229 [hep-ph]].
 
\bibitem{Gasser:1984ux}
  J.~Gasser and H.~Leutwyler,
  Nucl.\ Phys.\ B {\bf 250} (1985) 517.
 
\bibitem{Gasser:1984gg}
  J.~Gasser and H.~Leutwyler,
  Nucl.\ Phys.\ B {\bf 250} (1985) 465.

\bibitem{Lubicz:2010bv}
  V.~Lubicz {\it et al.}  [ETM Collaboration],
  PoS LATTICE {\bf 2010} (2010) 316
  [arXiv:1012.3573 [hep-lat]].

\bibitem{Ademollo:1964sr}
  M.~Ademollo and R.~Gatto,
  Phys.\ Rev.\ Lett.\  {\bf 13} (1964) 264.

\bibitem{Gasser:1986vb}
  J.~Gasser and H.~Leutwyler,
  Phys.\ Lett.\ B {\bf 184} (1987) 83.

\bibitem{Bernard:2006gy}
  V.~Bernard, M.~Oertel, E.~Passemar and J.~Stern,
  Phys.\ Lett.\ B {\bf 638} (2006) 480
  [hep-ph/0603202].
  
\bibitem{Bernard:2009zm}
  V.~Bernard, M.~Oertel, E.~Passemar and J.~Stern,
  Phys.\ Rev.\ D {\bf 80} (2009) 034034
  [arXiv:0903.1654 [hep-ph]].

\bibitem{Bazavov:2012cd}
  A.~Bazavov {\it et al.},
  Phys.\ Rev.\ D {\bf 87} (2013) 073012
  doi:10.1103/PhysRevD.87.073012
  [arXiv:1212.4993 [hep-lat]].

\bibitem{Boyle:2013gsa}
  P.~A.~Boyle, J.~M.~Flynn, N.~Garron, A.~Jüttner, C.~T.~Sachrajda, K.~Sivalingam and J.~M.~Zanotti,
  JHEP {\bf 1308} (2013) 132
  doi:10.1007/JHEP08(2013)132
  [arXiv:1305.7217 [hep-lat]].

\end{thebibliography}
\end{document}